\pdfoutput=1
\documentclass[letterpaper,titlepage,11pt]{article}

\usepackage{jheppub}
\usepackage[T1]{fontenc}
\usepackage[latin9]{inputenc}
\usepackage[active]{srcltx}
\usepackage{textcomp}
\usepackage{amsmath}
\usepackage{amssymb}
\usepackage{esint}
\usepackage{multirow}
\usepackage{empheq}
\usepackage{textcomp}

\usepackage{enumerate}
\usepackage{xcolor}
\def\be{\begin{eqnarray}}
\def\ee{\end{eqnarray}}
\def\beann{\begin{eqnarray*}}
\def\eeann{\end{eqnarray*}}
\def\beq{\begin{equation}}
\def\eeq{\end{equation}}
\def\ba{\begin{array}}
\def\ea{\end{array}}
\def\ben{\begin{enumerate}}
\def\een{\end{enumerate}}
\def\bea{\begin{eqnarray}}
\def\eea{\end{eqnarray}}

\providecommand{\Lt}{{\tt L}}
\renewcommand{\Lt}{{\tt L}}
\providecommand{\Wt}{{\tt W}}
\renewcommand{\Wt}{{\tt W}}

\providecommand{\Gt}{{\tt G}}
\renewcommand{\Gt}{{\tt G}}

\providecommand{\At}{{\tt A}}
\renewcommand{\At}{{\tt A}}
\providecommand{\St}{{\tt S}}
\renewcommand{\St}{{\tt S}}

\providecommand{\Jt}{{\tt J}}
\renewcommand{\Jt}{{\tt J}}

\def\be{\begin{equation}}
\def\ee{\end{equation}}
\def\bea{\begin{eqnarray}}
\def\eea{\end{eqnarray}}
\def\ba{\begin{array}}
\def\ea{\end{array}}
\def\nn{\nonumber}

\usepackage{amsfonts}

\usepackage{tikz}

\usepackage{bbm}

\usepackage{lineno}

\title{\bf{$\mathcal{N}=(2,2)$ extended $\mathfrak{sl}(3|2)$ Chern\,-\,Simons $AdS_3$ supergravity with new boundaries}}
\subheader{Prepared for submission to ArXiv.org  Preprint\\ \today}
\newcommand{\itu}{\dagger}

\author[\itu]{H. T. \"Ozer}
\emailAdd{ozert@itu.edu.tr}
\author[\itu]{,\,\,\,Ayt\"ul Filiz}
\emailAdd{aytulfiliz@itu.edu.tr}

\affiliation[\itu]{Istanbul Technical University,\,Faculty of Science and Letters,\,Physics Department,\\34469 Maslak,\,Istanbul,Turkey.}

\abstract{We present the first example of $\mathcal{N}=(2,2)$ formulation for the extended higher-spin $AdS_3$ supergravity with the most general boundary conditions as an extension of the $\mathcal{N}=(1,1)$ work, discovered recently by us \cite{Ozer:2019nkv}.\,Using the method proposed by Grumiller and Riegler,\,we construct a consistent class of the most general boundary conditions to extend it.\,An important consequence of our method is that,\,for the loosest set of boundary conditions it ensures that their asymptotic symmetry algebras consist of two copies of the $\mathfrak{sl}(3|2)_k$.\,Moreover,\,we enjoin some certain restrictions on the gauge fields for the most general boundary conditions,\,leading to the supersymmetric extensions of the Brown and Henneaux boundary conditions.\,Based on these results,\,we finally find out that the asymptotic symmetry algebras are two copies of the super $\mathcal{W}_3$ algebra for $\mathcal{N}=(2,2)$ extended higher\,-\,spin supergravity theory in $AdS_3$.
}

\makeatother
\begin{document}
\maketitle
\section{Introduction}
\hspace{0.5cm}
In modern theoretical physics, unquestionably,\,one of the forefront achievements in the past few decades is the discovery of the $AdS/CFT$ correspondence,\,which is first presented concretely by Maldacena \cite{Maldacena:1997re} in late 1997.\,This remarkable duality has profound implications ranging from a better understanding of many aspects of theoretical ( and even experimental) physics,\,especially general relativity,\,quantum gravity,\,quantum field theory,\,higher spin theory,\,and black holes.\,Moreover,\,the $AdS/CFT$ correspondence is an important manifestation of the holographic principle that posits a relation between a certain classical gravitational theory and a lower-dimensional non-gravitational one.\,The $AdS_3/CFT_2$ correspondence which is also a useful testing arena in this respect,\,implies an equivalence between pure Einstein $AdS_3$ gravity with a negative cosmological constant in 3D and a 2-dimensional conformal field theory.\,As a matter of fact, the main advantage of this eminent correspondence in three dimensions is to allow Einstein's gravity can be reformulated as a Chern\,-\,Simons gauge theory in such a way that all the structure is considerably simplified \cite{Achucarro:1987vz,Witten:1988hc}.\,What they discovered in their pioneer work is that,\,in three dimensions,\,the action and equations of motions are equivalent to a Chern\,-\,Simons  theory for an appropriate gauge group.\,

Despite the simplicity  owing to its topological nature,\,it is well\,-\,known that three dimensional gravity,\,besides being a very rich and spectacular theory since it has also been outstanding holographic properties.\,In this context,\,the striking feature of 3\,-\,dimensional  Einstein's Gravity is the absence of any local,\,propagating degrees of freedom,\,which means that any negatively curved Einstein space is locally $AdS_3$.\,There are only global degrees of freedom and hence no graviton in three dimensions.\,Notwithstanding that there are no local propagating degrees of freedom in the theory,\,its dynamic content is far from being insignificant due to the existence of boundary conditions.\,In other words,\,this means that the theory is wholly determined by global effects, since general relativity turns into a topological field theory, whose dynamics can be portrayed holographically by a 2\,-\,dimensional conformal field theory at the boundary.\,That is a Chern\,-\,Simons  theory in an equivalent formulation.\,At this point,\,it would be fair to say that the dynamics of the theory is totally presided over by boundary conditions.\,Furthermore,\,one should  call attention here that there is an infinite number of degrees of freedom living on the boundary under an appropriate choice of boundary conditions.\,These boundary conditions are requisite in order to provide that the action has a well-defined variational principle.\,Neverthless,\,their choice is not unique.\,Essentially,\,the dynamic features of the theory take shape according to the choice of these boundary conditions.\,So the residual gauge symmetry on the boundary within this framework emerges as global symmetry(asymptotic symmetry).\,

One of the most crucial results this story aforementioned tells us is that the asymptotic boundary conditions play a vital role in $AdS_3$ gravity.\,In their seminal paper \cite{Brown:1986nw},\,Brown and Hanneux proposed that under a convenient choice of boundary conditions,\,asymptotic symmetry algebras of $AdS_3$ gravity yields two copies of the Virasoro algebras with a classical central extension.\,The reason why this significant result is pointed out as a pioneer work is the fact that it is actually the first realization of $AdS/CFT$ correspondence,\,and also an important realization of holographic duality. Incidentally,\,another notable strength of $2+1$ dimensional gravity is that it also contains black hole solutions such as the famous BTZ black hole in which Einstein equations admit in the presence of the negative cosmological constant \cite{Banados:1992wn,Banados:1992gq}.\, Parenthetically,\,since the BTZ geometry is algebraically simple,\,it has provided quite a useful playground for studying features of black holes,\,which has a tremendous importance in exploring both classical and quantum gravitational physics.\,

It is worth mentioning that the Chern\,-\,Simons formulation of higher spin gravity in three dimensions has attracted more attention by the discovery of the Chern\,-\,Simons theories based on gauge algebras such as $\mathfrak{sl}(N|\mathbb{R})$) and $hs(\lambda)$ are versions of Vasiliev higher spin theories \cite{Vasiliev:1999ba,Vasiliev:2000rn} and also these are purely bosonic theories \cite{Bergshoeff:1989ns,Blencowe:1988gj} with higher spin fields of integer spin. Moreover, the Chern\,-\,Simons higher spin theories could pan out  with a realization of the classical $\mathcal{W}_N$ asymptotic symmetry algebras as in the related two-dimensional $CFT$'s \cite{Henneaux:2010xg,Campoleoni:2010zq,Campoleoni:2011hg,Campoleoni:2018uib,Tan:2011tj,Ozer:2017awd}.\,The promising results obtained from this perspective are adapted to extend the theory to the supergravity \cite{Brown:1986nw,Tan:2012xi,Banados:1998pi} as well as higher spin theory \cite{Blencowe:1988gj,Bergshoeff:1989ns}.\,Additionally,\,a supersymmetric generalization of these bosonic theories can be accomplished by keeping in view Chern\,-\,Simons theories based on superalgebras such as $\mathfrak{sl}(N|N-1)$, see,\,e.g.\, \cite{Candu:2012jq,Henneaux:2012ny,Tan:2012xi,Banados:1998pi,Peng:2012ae,Hanaki:2012yf},\,or $\mathfrak{osp}(N|N-1)$ \cite{Chen:2013oxa} which can be obtained by truncating out all the odd spin generators and one copy of the fermionic operators in $\mathfrak{sl}(N|N-1)$.

As already noted,\,one can casually state that three dimensional gravity is all about the choice of  boundary conditions.\,More precisely,\,the specification of boundary conditions is pivotal in comprehending how a theory that (locally) admits only a single solution.\,This analysis was first carried out by Brown and Henneaux \cite{Brown:1986nw} in their famous paper.\,Their study has also been encouraging to propose new sets of boundary conditions by sparking a vigorous research area which has gained in breadth over the years modifying \cite{Grumiller:2016pqb,Afshar:2015wjm,Compere:2013bya,Donnay:2015abr,Afshar:2016wfy,Troessaert:2013fma,Avery:2013dja,Perez:2016vqo} and generalizing \cite{Henneaux:2002wm,Henneaux:2004zi,Grumiller:2008es,Henneaux:2009pw,Oliva:2009ip,Barnich:2013sxa,Bunster:2014cna,Perez:2015jxn} these bc$'$s.\,In \cite{Grumiller:2016pqb}, Grumiller and Riegler have considered the most general $AdS_3$ boundary conditions, as a consequence,\,they have derived the asymptotic symmetry algebra consists of two $\mathfrak{sl}(2)_k$ current algebras.\,Furthermore,\,they have recovered all other previously found boundary conditions,\,imposing some certain restrictions to their most general boundary conditions.\,It is pertinent to address that there have been several papers recently inspired by them,\,i.e.,flat space \cite{Grumiller:2017sjh} and chiral higher spin gravity \cite{Krishnan:2017xct}, which is shown a new class of boundary conditions for higher spin theories in $AdS_3$.\,The simplest extension of Grumiller and Riegler's procedure for the most general $\mathcal{N}=(1,1)$,\,and $\mathcal{N}=(2,2)$ extended higher spin supergravity is introduced by Valcarcel \cite{Valcarcel:2018kwd} where the asymptotic symmetry algebra for the loosest set of boundary conditions for (extended) supergravity has been obtained.\,The most general $\mathcal{N}=(1,1)$ extended $AdS_3$ higher spin supergravity theory has been similarly presented, including further \cite{Ozer:2019nkv}.\,

This paper is concerned with the previously unresolved phenomenon;\,we construct a candidate solution for the most general $\mathcal{N}=(2,2)$ extended higher spin supergravity theory in $AdS_3$.\,We have shown that our theory falls under the same metric class as \cite{Grumiller:2017sjh},\,in which it was seen that the metric formulation could include even both charge and chemical potentials which are present in the Chern\,-\,Simons formalism.\,This can be considered as an alternative solution to the non\,-\,chiral Drinfeld\,-\,Sokolov type boundary conditions.\,Firstly,\,we have honed in on the simplest case $\mathcal{N}=(2,2)$ Chern\,-\,Simons theory based on the $\mathfrak{sl}(2|1)_k$ superalgebra.\,The related asymptotic symmetry algebra is two copies of the $\mathfrak{sl}(3|2)_k$ affine algebra.\,Then,\,we have tackled extended $\mathcal{N}=(2,2)$ Chern\,-\,Simons theory based on $\mathfrak{sl}(3|2)$ superalgebra,\,as a result,\,we have obtained asymptotic symmetry algebra consists of two copies of the $\mathfrak{sl}(3|2)_k$ affine algebra.\,Additionally,\,we have imposed some certain restrictions to the gauge fields on the most general boundary conditions,\,which leading us to the supersymmetric extensions of the Brown\,-\,Henneaux Boundary conditions.\,We have also shown that the asymptotic symmetry algebras are reduced to two copies of the super $\mathcal{W}_3$ algebra for the most general $\mathcal{N}=(2,2)$ extended higher spin $AdS_3$ supergravity theory.\,It is useful to indicate that it would be an interesting problem in its own right perform a different class of boundary conditions for (super)gravity that emerges in the literature (see,\,e.g.\,\cite{Compere:2013bya,Afshar:2016wfy,Troessaert:2013fma,Avery:2013dja}), since their higher spin generalization is not clear as  Grumiller and  Riegler's boundary conditions.\,In light of all these results,\,it is inevitable to say that this method yields an excellent laboratory to investigate the rich asymptotic structure of extended higher spin supergravity.\,

This paper is organized as follows.\,We first introduce a fundamental formulation of $\mathcal{N}=(2,2)$ supergravity as $\mathfrak{sl}(2|1) \oplus \mathfrak{sl}(2|1)$ Chern\,-\,Simons gauge theory for both affine and superconformal boundaries,\,respectively.\,Then,\,in section 3,\,we have maintained our calculations to extend the theory $\mathfrak{sl}(3|2) \oplus \mathfrak{sl}(3|2)$ higher\,-\,spin Chern\,-\,Simons supergravity in the case of both affine and superconformal boundaries,\,in where we putforthed explicitly principal embedding of $\mathfrak{sl}(2|1) \oplus \mathfrak{sl}(2|1)$ and also we came up with how asymptotic symmetry and higher spin Ward identities arise from these bulk equations of motion coupled to spin $s$,\,($s\,=\,1,\frac{3}{2},\frac{3}{2},2,2,\frac{5}{2},\frac{5}{2},3$) currents.\,We have dedicated this section to perform the asymptotic symmetry algebras as classical two copies of the $\mathfrak{sl}(3|2)_k$ affine algebra on the affine boundary and the super $\mathcal{W}_3$ symmetry algebra on the superconformal boundary,\,respectively.\,Besides,\,we have described the chemical potentials related to source fields appearing through the temporal components of the connection.\,In the final section,\,we conclude with a discussion,\,open issues and future research directions.\,

\section{Review of Chern\,-\,Simons supergravity in Three Dimensions:}\label{sec2}
In this section, we give a brief discussion for $AdS_3$ higher spin supergravity based on Chern\,-\,Simons formalism.
We especially employ this formulation to analyze $AdS_3$ supergravity in the presence of $\mathfrak{sl}(2|1)$ superalgebra
basis which has fallen into the same metric class as Grumiller and Riegler's recently proposed, the most general $AdS_3$ boundary conditions \cite{Grumiller:2016pqb}.\,
\subsection{Connection to Chern\,-\,Simons Theory}
In three dimensions, Einstein\,-\,Hilbert action for $\mathcal{N}=(2,2)$ supergravity with a negative cosmological constant,
can be defined in an equivalent Chern\,-\,Simons formulation over a spacetime manifold $\mathcal{M}$ as
\begin{equation}
S = S_{CS}[\Gamma] - S_{CS}[\bar{\Gamma}]\\
\end{equation}
where
\begin{equation}
S_{CS}[\Gamma] = \frac{k}{4\pi}\int_{\mathcal{M}} \mathfrak{str}\bigg(\Gamma\wedge \mathrm{d}\Gamma\,+\,\frac{2}{3}\,\Gamma\wedge\Gamma\wedge\Gamma \bigg)
\end{equation}
which was first noted by Achucarro and Townsend \cite{Achucarro:1987vz} and further developed by Witten \cite{Witten:1988hc}.

The Chern\,-\,Simons level $k$ will eventually be related to ratio of $AdS_3$ radius $l$ and Newton constant $G$, and also the related  central charge $c$ of
the superconformal field theory as\\$k=\frac{\ell}{8G\mathfrak{str}( {\tt L}_{0} {\tt L}_{0}) }=\frac{c}{12\mathfrak{str}( {\tt L}_{0} {\tt L}_{0}) }$. Notice that while the 1\,-\,forms $(\Gamma,\bar{\Gamma})$ connections are defined as to take values in the gauge group of $\mathfrak{sl}(2|1)$ superalgebra, the supertrace  $\mathfrak{str}$ which shows a metric on the $\mathfrak{sl}(2|1)$ Lie superalgebra, is taken over the superalgebra generators.

It is convenient to get started with standard basis for $\mathfrak{sl}(2|1)$ Lie superalgebra. We have denoted the bosonic generators by\,$\Lt_{i}$ $(i=\pm1,0)$, $\Jt$ and the fermionic ones by \,$\Gt_{r}^{M}$ $(r=\pm\frac{1}{2},\,M=\pm)$, whose commutations relations read
\begin{eqnarray}
\label{algebraLG}
\left[\Lt_{i},\Lt_{j}\right]&=& \left(i-j\right)\Lt_{i+j},~~
\left[\Lt_{i},\Gt_{r}^{\pm}\right]=\left(\frac{i}{2}-r\right)\Gt_{i+r}^{\pm},~~
\left[\Jt,\Gt_{r}^{\pm}\right]=\pm\Gt_{r}^{\pm},\,\\
\left\{\Gt_{r}^{\pm},\Gt_{s}^{\mp}\right\} &=&2\,\Lt_{r+s}\pm\left(r-s\right) \Jt
\end{eqnarray}
except for zero commutators.

The Chern\,-\,Simons equations of motions, also known as the flatness conditions correspond to vanishing field strengths; $F = \bar{F} =0$ where
\begin{equation}
\label{flatness}
F= \mathrm{d}\Gamma + \Gamma\wedge \Gamma =0,\,\,\,\,\,\,\bar{F}= \mathrm{d}\bar{\Gamma} + \bar{\Gamma}\wedge \bar{\Gamma} =0
\end{equation}
which is equivalent to Einstein's equation. The relation to the Einstein's equation is made by expressing Lie algebra valued generalizations of the vielbein and spin connection in terms of the gauge connections. Then, one can obtain the metric $g_{\mu \nu}$ from the vielbein $e = \frac{\ell}{2}(\Gamma- \bar{\Gamma})$ in the usual fashion
\begin{equation}\label{metric}
g_{\mu \nu} =  \frac{1}{2}\,\mathfrak{str}( e_{\mu} e_{\nu} ).
\end{equation}
By the choice of the radial gauge, asymptotically $AdS_3$ connections can be taken to have the form
\begin{eqnarray}
\label{ads31}
\Gamma&=& b^{-1} a\left(t,\phi\right) b +b^{-1} db,\,\,\,\,\,\,  \bar{\Gamma}=b\bar{a}\left(t,\phi\right)b^{-1} + bdb^{-1}
\end{eqnarray}
with state\,-\,independent group element (called Grumiller-Riegler gauge);
\begin{equation}
b=e^{\Lt_{-1}}e^{\rho \Lt_{0}}
\end{equation}
which yields a more general metric and means that it includes all $\mathfrak{sl}(2|1)$ charges and chemical potentials can be chosen accordingly. At this point, it is important to note that as long as $\delta b=0$, the choice of $b$ is irrelevant for asymptotic symmetries.
Unlike the standard choice of $b$, this freedom enables a more general metric. Therefore, it is cruical to choose the most general boundary conditions preserving this most general metric form for supergravity.

Further,\,in the radial gauge $a\left(t,\phi\right)$ and $\bar{a}\left(t,\phi\right)$ connections are the $\mathfrak{sl}(2|1)$ Lie superalgebra valued fields which have been independent of a radial coordinate as
\begin{eqnarray}\label{conn}
a\left(t,\varphi\right) = a_{t}\left(t,\varphi\right)\mathrm{d}t+a_{\varphi}\left(t,\varphi\right)\mathrm{d}\varphi\nonumber\\
\bar{a}\left(t,\varphi\right) = \bar{a}_{t}\left(t,\varphi\right)\mathrm{d}t+\bar{a}_{\varphi}\left(t,\varphi\right)\mathrm{d}\varphi
\end{eqnarray}
Hereafter only focused on the unbarred sector, since the analysis of the barred sector works in complete analogy yielding the same outcomes with the barred sector and it can be figured out by the same algorithm thanks to the procedure used.
\subsection{$\mathfrak{sl}(2|1) \oplus \mathfrak{sl}(2|1)$ Chern\,-\,Simons $\mathcal{N}=(2,2)$ Supergravity  for Affine Boundary}\label{osp21}
We begin by reviewing asymptotically $AdS_3$ boundary conditions for $\mathfrak{sl}(2|1) \oplus \mathfrak{sl}(2|1)$ Chern\,-\,Simons theory in the affine case. We present how the procedure mentioned in \cite{Grumiller:2016pqb} can be used to evaluate the asymptotic symmetry algebra. According to the results obtained, the most general solution of Einstein's equation that is asymptotically $AdS_3$ is defined by the following general metric form:
\begin{eqnarray}\label{sugra08}
    \mathrm{d}s^2 &=& \mathrm{d}\rho^2 + 2\left[ e^\rho N^{(0)}_i +  N^{(1)}_i +  e^{-\rho} N^{(2)}_i + \mathcal O \left( e^{-2\rho}\right)\right]\mathrm{d}\rho \mathrm{d}x^i\nonumber\\
    &+& \left[ e^{2\rho} g^{(0)}_{ij} +  e^\rho g^{(1)}_{ij} +  g^{(2)}_{ij} + \mathcal O \left( e^{-\rho}\right)\right]\mathrm{d}x^i \mathrm{d}x^j.
\end{eqnarray}
So, it is important to define the most general $\mathcal{N}=(2,2)$ supergravity boundary conditions which preserve this form of the metric.

We start by proposing $\mathfrak{sl}(2|1)$ Lie superalgebra valued $a_\varphi$ component of the gauge connection in the form:
\begin{eqnarray}\label{bouncond999}
a_\varphi &=&\rho\mathcal{J}\Jt +\gamma_i\mathcal{L}^i \Lt_{i}+\sigma_M^p\mathcal{G}_{M}^{p}\Gt_{p}^{M}
\end{eqnarray}
where $\rho=\frac{1}{k}$,\,$\frac{\gamma_{0}}{2}=-\gamma_{\pm1}=\frac{2}{k}$,\,$\sigma_{\pm}^{-\frac{1}{2}}=-\sigma_{\pm}^{\frac{1}{2}}=\frac{1}{k}$
are some scaling parameters to be identified later. We have eight state-dependent functions consisting of four bosonic $\left(\mathcal{J},\mathcal{L}^i\right)$ and four fermionic $\mathcal{G}_{M}^{p}$, usually called as $charges$. The time component $a_t$ of the connection $a\left(t,\varphi\right)$ can be given as
\begin{eqnarray}\label{bouncond888}
a_t &=&\eta\Jt +\mu^i \Lt_{i}+\nu_{M}^{p}\Gt_{p}^{M}.
\end{eqnarray}
In this case, we have eight independent functions $(\eta,\mathcal{\mu}^i,\nu_{M}^{p})$, as chemical potentials which are not allowed to vary, $\delta a_t= 0$.

Using the flatness conditions $(\ref{flatness})$, the equations of motions for fixed chemical potentials impose the following additional
conditions on the charges $(\mathcal{J},\mathcal{L}^i,\mathcal{G}_{M}^{p})$:
\newpage
\begin{eqnarray}
2\partial_{t}\mathcal{L}^{0}&=&
\frac{k}{2}\partial_{\varphi}\mu^{0}
+2\mathcal{L}^{+1}\mu^{-1}
+2\mathcal{L}^{-1}\mu^{+1}
-\mathcal{G}_{-}^{+\frac{1}{2}}\nu_{+}^{-\frac{1}{2}}
+\mathcal{G}_{-}^{-\frac{1}{2}}\nu_{+}^{+\frac{1}{2}}
-\mathcal{G}_{+}^{+\frac{1}{2}}\nu_{-}^{-\frac{1}{2}}
+\mathcal{G}_{+}^{-\frac{1}{2}}\nu_{-}^{+\frac{1}{2}},\\
\partial_{t}\mathcal{L}^{\pm 1}	&=&	
-\frac{k}{2}\partial_{\varphi}\mu^{\pm 1}
\pm  \mathcal{L}^{0}\mu^{\pm 1}
\pm   \mathcal{L}^{\pm 1}\mu^{0}
+\mathcal{G}_{+}^{\pm\frac{1}{2}}\nu_{-}^{\pm\frac{1}{2}}
+\mathcal{G}_{-}^{\pm\frac{1}{2}}\nu_{+}^{\pm\frac{1}{2}},  \\
\partial_{t}\mathcal{G}_{\pm}^{\pm\frac{1}{2}}&=&	
\pm k\partial_{\varphi}\nu_{\pm}^{\pm\frac{1}{2}}
\pm 2\mathcal{L}^{\pm}\nu_{\pm}^{\mp\frac{1}{2}}
+2\mathcal{L}^{0}\nu_{\pm}^{\pm\frac{1}{2}}
\pm \mathcal{G}_{\pm}^{\mp\frac{1}{2}}\mu^{\pm}
\pm\frac{1}{2} \mu^{0}\mathcal{G}_{\pm}^{\pm\frac{1}{2}}
\mp\mathcal{J}\nu_{\pm}^{\pm\frac{1}{2}}
-\eta \mathcal{G}_{\pm}^{\pm\frac{1}{2}}, \\
\partial_{t}\mathcal{J}	&=&
k\partial_{\varphi}\eta
+\mathcal{G}_{-}^{+\frac{1}{2}}\nu_{+}^{-\frac{1}{2}}
+\mathcal{G}_{-}^{-\frac{1}{2}}\nu_{+}^{+\frac{1}{2}}
-\mathcal{G}_{+}^{+\frac{1}{2}}\nu_{-}^{-\frac{1}{2}}
-\mathcal{G}_{+}^{-\frac{1}{2}}\nu_{-}^{+\frac{1}{2}},
\end{eqnarray}
that represents the temporal evolution of the eight state-dependent source fields.

We want to derive asymptotic symmetry algebra for the most general boundary conditions through a canonical analysis. That's why we embark on by considering all gauge transformations:
\begin{equation}\label{boundarycond}
    \delta_{\lambda}\Gamma = \mathrm{d}\lambda + \left[\Gamma,\lambda \right]
\end{equation}
which preserve the most general boundary conditions. At this point, it would be appropriate to single out the gauge parameter in terms of the $\mathfrak{sl}(2|1)$ Lie superalgebra basis
 \begin{equation}
    \lambda=b^{-1}\bigg[
  \varrho\Jt
 +\epsilon^i \Lt_{i}
 +\zeta_{M}^{p}\Gt_{p}^{M}
    \bigg]b\label{boundarycond2}.
\end{equation}
Note that the gauge parameter includes four bosonic $\varrho$,\,$\mathcal{\epsilon}^i$ and four fermionic $\mathcal{\zeta}_{M}^{p}$, arbitrary functions of boundary coordinates. And also, we are concerned with the gauge parameters that satisfy $(\ref{boundarycond})$. One can now determine the boundary preserving gauge transformations.
Accordingly, the infinitesimal gauge transformations are given by;
\begin{eqnarray}
2\partial_{t}\mathcal{L}^{0}&=&
\frac{k}{2}\partial_{\varphi}\epsilon^{0}
+2\mathcal{L}^{+1}\epsilon^{-1}
+2\mathcal{L}^{-1}\epsilon^{+1}
-\mathcal{G}_{-}^{+\frac{1}{2}}\zeta_{+}^{-\frac{1}{2}}
+\mathcal{G}_{-}^{-\frac{1}{2}}\zeta_{+}^{+\frac{1}{2}}
-\mathcal{G}_{+}^{+\frac{1}{2}}\zeta_{-}^{-\frac{1}{2}}
+\mathcal{G}_{+}^{-\frac{1}{2}}\zeta_{-}^{+\frac{1}{2}},\label{evaa41} \\
\partial_{t}\mathcal{L}^{\pm 1}	&=&	
-\frac{k}{2}\partial_{\varphi}\epsilon^{\pm 1}
\pm  \mathcal{L}^{0}\epsilon^{\pm 1}
\pm   \mathcal{L}^{\pm 1}\epsilon^{0}
+\mathcal{G}_{+}^{\pm\frac{1}{2}}\zeta_{-}^{\pm\frac{1}{2}}
+\mathcal{G}_{-}^{\pm\frac{1}{2}}\zeta_{+}^{\pm\frac{1}{2}}, \\
\partial_{t}\mathcal{G}_{\pm}^{\pm\frac{1}{2}}&=&	
\pm k\partial_{\varphi}\zeta_{\pm}^{\pm\frac{1}{2}}
\pm 2\mathcal{L}^{\pm}\zeta_{\pm}^{\mp\frac{1}{2}}
+2\mathcal{L}^{0}\zeta_{\pm}^{\pm\frac{1}{2}}
\pm \mathcal{G}_{\pm}^{\mp\frac{1}{2}}\epsilon^{\pm}
\pm\frac{1}{2} \epsilon^{0}\mathcal{G}_{\pm}^{\pm\frac{1}{2}}
\mp\mathcal{J}\zeta_{\pm}^{\pm\frac{1}{2}}
-\varrho \mathcal{G}_{\pm}^{\pm\frac{1}{2}},  \\
\partial_{t}\mathcal{J}	&=&
k\partial_{\varphi}\varrho
+\mathcal{G}_{-}^{+\frac{1}{2}}\zeta_{+}^{-\frac{1}{2}}
+\mathcal{G}_{-}^{-\frac{1}{2}}\zeta_{+}^{+\frac{1}{2}}
-\mathcal{G}_{+}^{+\frac{1}{2}}\zeta_{-}^{-\frac{1}{2}}
-\mathcal{G}_{+}^{-\frac{1}{2}}\zeta_{-}^{+\frac{1}{2}}.\label{evaa51}
\end{eqnarray}
One can also derive the following constraints for the chemical potentials analogously.
\begin{eqnarray}
2\partial_{t}\mu^{0}&=&
\frac{k}{2}\partial_{\varphi}\epsilon^{0}
+2\mu^{+1}\epsilon^{-1}
+2\mu^{-1}\epsilon^{+1}
-\nu_{-}^{+\frac{1}{2}}\zeta_{+}^{-\frac{1}{2}}
+\nu_{-}^{-\frac{1}{2}}\zeta_{+}^{+\frac{1}{2}}
-\nu_{+}^{+\frac{1}{2}}\zeta_{-}^{-\frac{1}{2}}
+\nu_{+}^{-\frac{1}{2}}\zeta_{-}^{+\frac{1}{2}},\\
\partial_{t}\mu^{\pm 1}	&=&	
-\frac{k}{2}\partial_{\varphi}\epsilon^{\pm 1}
\pm  \mu^{0}\epsilon^{\pm 1}
\pm   \mu^{\pm 1}\epsilon^{0}
+\nu_{+}^{\pm\frac{1}{2}}\zeta_{-}^{\pm\frac{1}{2}}
+\nu_{-}^{\pm\frac{1}{2}}\zeta_{+}^{\pm\frac{1}{2}}, \\
\partial_{t}\nu_{\pm}^{\pm\frac{1}{2}}&=&	
\pm k\partial_{\varphi}\zeta_{\pm}^{\pm\frac{1}{2}}
\pm 2\mu^{\pm}\zeta_{\pm}^{\mp\frac{1}{2}}
+2\mu^{0}\zeta_{\pm}^{\pm\frac{1}{2}}
\pm \nu_{\pm}^{\mp\frac{1}{2}}\epsilon^{\pm}
\pm\frac{1}{2} \epsilon^{0}\nu_{\pm}^{\pm\frac{1}{2}}
\mp\eta\zeta_{\pm}^{\pm\frac{1}{2}}
-\varrho \nu_{\pm}^{\pm\frac{1}{2}}, \\
\partial_{t}\eta	&=&
k\partial_{\varphi}\varrho
+\nu_{-}^{+\frac{1}{2}}\zeta_{+}^{-\frac{1}{2}}
+\nu_{-}^{-\frac{1}{2}}\zeta_{+}^{+\frac{1}{2}}
-\nu_{+}^{+\frac{1}{2}}\zeta_{-}^{-\frac{1}{2}}
-\nu_{+}^{-\frac{1}{2}}\zeta_{-}^{+\frac{1}{2}}.
\end{eqnarray}
As a final step, the canonical boundary charge $\mathcal{Q[\lambda]}$ that generates the transformations $(\ref{evaa41})$-$(\ref{evaa51})$ can be defined. For this purpose, the variation of the canonical boundary charge $\mathcal{Q[\lambda]}$ \cite{Banados:1998gg,Carlip:2005zn,Banados:1994tn,Banados:1998ta} leading the asymptotic symmetry algebra is given by
\begin{equation}\label{Qvar}
\delta_\lambda \mathcal{Q}=\frac{k}{2\pi}\int\mathrm{d}\varphi\;\mathfrak{str}\left(\lambda \delta \Gamma_{\varphi}\right).
\end{equation}
Hence, the variation of the canonical boundary charge $\mathcal{Q[\lambda]}$ can be functionally integrated to yield
\begin{equation}\label{boundaryco10}
\mathcal{Q[\lambda]}=\int\mathrm{d}\varphi\;\left[\mathcal{J} \varrho + \mathcal{L}^{i}\epsilon^{-i}
+\mathcal{G}_{M}^{p}\zeta_{M}^{-p}\right].
\end{equation}
After both having determined the infinitesimal transformations and the canonical boundary charge, now we are in a position to derive the asymptotic symmetry algebra using the standart method \cite{Blgojevic:2002}, which can be obtained through the following relation
\begin{equation}\label{Qvar2}
\delta _{\lambda}\digamma\,=\,\{\digamma,\mathcal{Q}[\lambda]\}
\end{equation}
for any phase space functional $\digamma$. The Poisson brackets of all fields can be calculated as
\begin{eqnarray}\label{bracketalgeb}
\{\mathcal{L}^i(z_1),\mathcal{L}^j(z_2)\}_{_{PB}}\,&=&\,(i-j) \mathcal{L}^{i+j}(z_2)\delta(z_1-z_2)-k\eta_{2}^{ij}\partial_{\varphi}\delta(z_1-z_2),\\
\{\mathcal{L}^i(z_1),\mathcal{G}_{\pm}^p(z_2)\}_{_{PB}}\,&=&\,\left(\frac{i}{2}-p\right) \mathcal{G}_{\pm}^{i+p}(z_2)\delta(z_1-z_2),\\
\{\mathcal{J}(z_1),\mathcal{G}_{\pm}^p(z_2)\}_{_{PB}}\,&=&\,\pm\mathcal{G}_{\pm}^p(z_2)\delta(z_1-z_2),\\
\{\mathcal{J}(z_1),\mathcal{J}(z_2)\}_{_{PB}}\,&=&\,-k\eta\partial_{\varphi}\delta(z_1-z_2),\\
\{\mathcal{G}_{\pm}^p(z_1),\mathcal{G}_{\pm}^q(z_2)\}_{_{PB}}\,&=& \bigg(2\mathcal{L}^{p+q}(z_2)\pm(p-q) \mathcal{J}\bigg)\delta(z_1-z_2)\nonumber\\
                                                           &+&k\eta_{\frac{3}{2}}^{pq}\partial_{\varphi}\delta(z_1-z_2).
\end{eqnarray}
where $\eta=\mathfrak{str}(\Jt\Jt)$,\,\,$\eta_{2}^{ij}=\mathfrak{str}(\Lt_i\Lt_j)$\,and \,$\eta_{\frac{3}{2}}^{pq}=\mathfrak{str}(\Gt_p^{M}\Gt_q^{M})$
are the bilinear forms in the fundamental representation of $\mathfrak{sl}(2|1)$ Lie superalgebra. $\mathcal{J}(z)$  and $\mathcal{G}_{M}^{p}(z)$ charges are also identified as the generators of the asymptotic symmetry algebra. Finally, the operator product algebra can be written as
\begin{eqnarray}\label{ope22}
\mathcal{L}^{i}(z_1)\mathcal{L}^{j}(z_2)\,& \sim &\, \frac{\frac{k}{2}\eta_{2}^{ij}}{z_{12}^{2}}\,+\,\frac{(i-j)}{z_{12}} \mathcal{L}^{i+j}(z_2),\\
\mathcal{L}^{i}(z_1)\mathcal{G}_{\pm}^{p}(z_2)\,& \sim &\,\frac{({i\over 2}-p)}{z_{12}} \mathcal{G}_{\pm}^{i+p}(z_2),\\
\mathcal{J}(z_1)\mathcal{G}_{\pm}^{p}(z_2)\,& \sim &\,\pm\frac{\mathcal{G}_{\pm}^p(z_2)}{z_{12}},\\
\mathcal{J}(z_1)\mathcal{J}(z_2)\,& \sim &  \frac{\frac{k}{2}\eta}{z_{12}^2},\\
\mathcal{G}_{\pm}^{p}(z_1)\mathcal{G}_{\pm}^{q}(z_2)\,& \sim &\, \frac{\frac{k}{2}\eta_{\frac{3}{2}}^{pq}}{z_{12}^{2}}\,+\,\frac{2}{z_{12}}\bigg( \mathcal{L}^{p+q}(z_2)\pm\frac{(p-q)}{2} \mathcal{J}(z_2)\bigg).
\end{eqnarray}
where $z_{12}=z_1-z_2$,\,or in the more compact form,
\begin{eqnarray}\label{ope11}
\mathfrak{\mathcal{\mathfrak{J}}}^{A}(z_1)\mathfrak{\mathcal{\mathfrak{J}}}^{B}(z_2)\,& \sim &\, \frac{\frac{k}{2}\eta^{AB}}{z_{12}^{2}}\,+\,\frac{\mathfrak{\mathcal{\mathfrak{f}}}^{AB}_{~~~C} \mathfrak{\mathcal{\mathfrak{J}}}^{C}(z_2)}{z_{12}}.
\end{eqnarray}
Note that $\eta^{AB}$ is the supertrace matrix and $\mathfrak{\mathcal{\mathfrak{f}}}^{AB}_{~~C}$'s are the structure constants of the related algebra with $(A,B=0,\pm1,\pm\frac{1}{2})$,\,i.e,\,$\eta^{ip}=0$ and $\mathfrak{\mathcal{\mathfrak{f}}}^{ij}_{~~i+j}=(i-j)$. Lastly, by repeating the same analysis for the barred sector also, the asymptotic symmetry algebra of $\mathcal{N}=(2,2)$ supergravity for the loosest set of boundary conditions is given by two copies of the affine $\mathfrak{sl}(2|1)_k$ algebra.
\subsection{$\mathfrak{sl}(2|1) \oplus \mathfrak{sl}(2|1)$  Chern\,-\,Simons $\mathcal{N}=(2,2)$ Supergravity for Superconformal Boundary}\label{bhreduction1}
In this section, our aim is to look into the asymptotic symmetry algebra for the supersymmetric extension of the Brown-Henneaux boundary conditions.
We start on by imposing the Drinfeld-Sokolov heighest weight gauge condition on the $\mathfrak{sl}(2|1)$ Lie superalgebra valued connection $(\ref{bouncond999})$, in order to further restrict the coefficients. So, the Drinfeld-Sokolov reduction sets the fields such that
\begin{eqnarray}
\mathcal{L}^0=\mathcal{G}_{\pm}^{+\frac{1}{2}}=0,\,\,\,\,\,\,\,\,\,\mathcal{L}^{-1}=\mathcal{L},\,\,\,\,\,\,\,\,\,\mathcal{G}_{\pm}^{-\frac{1}{2}}=\mathcal{G}_{\pm},\,\,\,\,\,\,\,\,\,\gamma_{+1}\mathcal{L}^{+1}=1.
\end{eqnarray}
Meanwhile, it is worth noting that the super-conformal boundary conditions are the supersymmetric extension of the well-known Brown-Henneaux boundary conditions proposed in \cite{Brown:1986nw} for $AdS_3$ supergravity. As a result, the supersymmetric gauge connection takes the form
\begin{eqnarray}\label{bouncondhf}
a_\varphi &=&\Lt_{+1} +\gamma_{-1}\mathcal{L} \Lt_{-1}+\sigma_{\pm}^{-{\frac{1}{2}}}\mathcal{G}_{\pm}\Gt_{-{\frac{1}{2}}}^{\pm}+\rho\mathcal{J}\Jt,
\end{eqnarray}
where $\gamma_{-1}=-\frac{1}{k}$,\,$\rho=\frac{1}{2k}$,\,$\sigma_{+}^{-{\frac{1}{2}}}=\frac{1}{2k}$,\,and $\sigma_{-}^{-{\frac{1}{2}}}=-\frac{1}{2k}$'s are some scaling parameters and we have four functions:\,two bosonic $\left(\mathcal{J},\mathcal{L}\right)$ and two fermionic $\mathcal{G}_{\pm}$ as $charges$. After performing these steps, we are now close to acquire the superconformal asymptotic symmetry algebra. Following the results implied by the Drinfeld-Sokolov reduction, the gauge parameter $\lambda$ has only four independent functions $(\varrho,\mathcal{\epsilon}\equiv\epsilon^{+1},\zeta_{\pm}\equiv \zeta_{\pm}^{+\frac{1}{2}})$ and given as
\begin{eqnarray}\label{ebc08}
\lambda &=& b^{-1}\left[
\epsilon \Lt_{1}
-\epsilon'\Lt_{0}
+\left(
       \frac{1}{2}\epsilon''
      +\frac{1}{k}\mathcal{L}\epsilon
      +\frac{1}{2k}\mathcal{G}_{+}\zeta_{+}
      +\frac{1}{2k}\mathcal{G}_{-}\zeta_{-}
 \right)\Lt_{-1}
+\zeta_{+}\Gt_{\frac{1}{2}}^{+}
+\zeta_{-}\Gt_{\frac{1}{2}}^{-}
\right.\nonumber\\
&&\left.
-\left(\frac{1}{2k}\mathcal{G}_{-}\epsilon+\frac{1}{2k}\mathcal{J}\zeta_{+}+\zeta_{+}'\right)\Gt_{-\frac{1}{2}}^{-}
-\left(\frac{1}{2k}\mathcal{G}_{+}\epsilon+\frac{1}{2k}\mathcal{J}\zeta_{-}+\zeta_{-}'\right)\Gt_{-\frac{1}{2}}^{+}
+\varrho\Jt
\right]b.
\end{eqnarray}
Substituting this gauge parameter in the transformation of the fields expression $(\ref{boundarycond})$, we obtain the infinitesimal gauge transformations:
\begin{eqnarray}
\delta_{\lambda}\mathcal{J}&=&\frac{1}{2k}\varrho'-\mathcal{G}_{+}\zeta_{+}+\mathcal{G}_{-}\zeta_{-},\\
\delta_{\lambda}\mathcal{L}&=&
\frac{k}{2}\epsilon'''
+2\mathcal{L}\epsilon'
+\mathcal{L}'\epsilon
+\frac{3}{2}\mathcal{G}_{+}\zeta_{+}'
+\frac{3}{2}\mathcal{G}_{-}\zeta_{-}'
+\frac{1}{2}\mathcal{G}_{+}'\zeta_{+}
+\frac{1}{2}\mathcal{G}_{-}'\zeta_{-}\nonumber
\end{eqnarray}
\begin{eqnarray}
&+&\frac{1}{2k}\big(\mathcal{J}\mathcal{G}_{+}\zeta_{+}+\mathcal{J}\mathcal{G}_{-}\zeta_{-}\big),\\
\delta_{\lambda}\mathcal{G}_{\pm}&=&\frac{1}{2k}\zeta_{\mp}''
+2\bigg(\mathcal{L}+\frac{1}{4k}\big(\mathcal{J}\mathcal{J}\big)\bigg)\zeta_{\pm}
\pm\left(\varrho-\frac{1}{2k}\mathcal{J}\epsilon\right)\mathcal{G}_{\pm}
\mp\zeta_{\pm}\mathcal{J}'
+\frac{3}{2}\mathcal{G}_{\pm}\epsilon'\nonumber\\
&+&\epsilon \mathcal{G}_{\pm}'
\mp\mathcal{J}\zeta_{\mp}'.
\end{eqnarray}
In fact, these gauge transformations give a cue for the asymptotic symmetry algebra \cite{Banados:1998pi}. By taking forward to see the asymmptotic symmetries, one can integrate the variation of the canonical boundary charges, i.e., $\delta_\lambda \mathcal{Q}$ expression $(\ref{Qvar})$ such that
\begin{equation}
\mathcal{Q[\lambda]}=\int\mathrm{d}\varphi\;\left[\mathcal{L}\epsilon+\mathcal{G}_M\zeta_M +\mathcal{J}\varrho\right].\label{boundaryco11111333333}
\end{equation}
But,\,these canonical boundary charges do not give a convenient asymptotic operator product algebra for $\mathcal{N}=(2,2)$ superconformal boundary,
\begin{eqnarray}\label{ope2}
\mathcal{J}(z_1)\mathcal{J}(z_2)\,&\sim&  \,{{2k}\over{z_{12}^{2}}},~~~\mathcal{J}(z_1)\mathcal{G}_{\pm}(z_2)\, \sim  \,{{\mp\mathcal{G}_{\pm}}\over{z_{12}^{2}}},\,\\
\mathcal{L}(z_1)\mathcal{L}(z_2)\,&\sim & \,
{{3k}\over{z_{12}^{4}}}\,
+\, {2\,\mathcal{L}\over{z_{12}^{2}}}\,
+ \,{\mathcal{L}'-\frac{\mathcal{G}_{+}\mathcal{G}_{-}}{k}\over{z_{12}}},~~~\mathcal{L}(z_1)\mathcal{J}(z_2)\,\sim\,\,0 \\
\mathcal{L}(z_1)\mathcal{G}_{\pm}(z_2)\, &\sim & \,{{\frac{3}{2}\mathcal{G}_{\pm}}\over{z_{12}^{2}}}\, + \,\frac{\mathcal{G}_{\pm}'\pm\frac{\mathcal{J}\mathcal{G}_{\pm}}{2k}}{z_{12}},\\
\mathcal{G}_{\pm}(z_1)\mathcal{G}_{\pm}(z_2)\,&\sim & \,{\mp{4k}\over{z_{12}^{3}}}\,-\,{{2\mathcal{J}}\over{z_{12}^{2}}}\,\, + \,\frac{\mp2}{z_{12}}\big( \mathcal{L}+\frac{\mathcal{J}\mathcal{J}}{4k}\pm\frac{\mathcal{J}'}{2}\big).
\end{eqnarray}
 because $\mathcal{L}(z)$ and $\mathcal{J}(z)$ do not transform like a primary conformal field, besides there exist some nonlinear terms such as $(\mathcal{JJ})(z)$,\,$(\mathcal{G}_{+}\mathcal{G}_{-})(z)$ and also $(\mathcal{J}\mathcal{G}_{\pm})(z)$. Therefore, it is neccessary to perform a shift on the boundary charge $\mathcal{L}$ and also make a redefinition on the gauge parameter $\varrho$ as follows:
\begin{equation}\label{shift1}
\mathcal{L}\rightarrow\mathcal{L}+\frac{3}{2c} \big(\mathcal{J}\mathcal{J}\big),\,~~~~\varrho\rightarrow\varrho+\frac{3}{c}\mathcal{J}\epsilon.
\end{equation}
Before closing this section, one should also emphasize that these new variables do not affect the boundary charges. Thus, this leads to operator product expansions of the convenient asymptotic symmetry algebra for $\mathcal{N}=(2,2)$ superconformal boundary with a set of conformal generators $\mathcal{G}_{\pm}\rightarrow\mathcal{G}^{+}\pm\mathcal{G}^{-}$ in the complex coordinates by using $(\ref{Qvar2})$
\begin{eqnarray}\label{ope2}
&\mathcal{J}(z_1)\mathcal{J}(z_2)\,\sim  \,{{c\over 3}\over{z_{12}^{2}}},~~~\mathcal{J}(z_1)\mathcal{G}^{\pm}(z_2)\, \sim  \,{{\pm\mathcal{G}^{\pm}}\over{z_{12}^{2}}},\,\\
&\mathcal{L}(z_1)\mathcal{L}(z_2)\,\sim  \,{{c\over 2}\over{z_{12}^{4}}}\,+\, {2\,\mathcal{L}\over{z_{12}^{2}}}\, + \,{ \mathcal{L}'\over{z_{12}}},~~~\mathcal{L}(z_1)\mathcal{J}(z_2)\,\sim\,{\,\mathcal{J}\over{z_{12}^{2}}}\, + \,{ \mathcal{J}'\over{z_{12}}} \\
&\mathcal{L}(z_1)\mathcal{G}^{\pm}(z_2)\, \sim  \,{{\frac{3}{2}\mathcal{G}^{\pm}}\over{z_{12}^{2}}}\, + \,\frac{\mathcal{G}^{\pm'}}{z_{12}},\\
&\mathcal{G}^{\pm}(z_1)\mathcal{G}^{\mp}(z_2)\,\sim \,{{2c\over 3}\over{z_{12}^{3}}}\,+\,{{2\mathcal{J}}\over{z_{12}^{2}}}\,\, + \,\frac{1}{z_{12}}\big(2\mathcal{L}\pm\mathcal{J}'\big).
\end{eqnarray}
When the same analysis has repeated for the barred sector, it is seen that the asymptotic symmetry algebra for the loosest set of boundary conditions of $\mathcal{N}=(2,2)$ supergravity, consists of two copies of the super-Virasoro algebra with central charge $c\,=\,6k$.
\section{$\mathcal{N}=(2,2)$ $\mathfrak{sl}(3|2) \oplus \mathfrak{sl}(3|2)$ Higher\,-\,Spin  Chern\,-\,Simons Supergravity}
Finally, after having laid the groundwork by performing the canonical analysis of the simplest case, now we are in a position to present the extended $\mathcal{N}=(2,2)$ higher\,-\,spin Chern\,-\,Simons supergravity theory based on $\mathfrak{sl}(3|2)_k$ superalgebra.
\subsection{For Affine Boundary}
The objective of this section is to construct $\mathcal{N}=(2,2)$ extended higher-spin $AdS_3$ supergravity as $\mathfrak{sl}(3|2)\oplus\mathfrak{sl}(3|2)$
Chern\,-\,Simons gauge theory on the affine boundary. We proceed our calculations to elucidate the asymptotic symmetry algebra for the loosest set of boundary conditions.

As already stated in the previous section, we consider the principal embedding of $\mathfrak{sl}(2|1)$ into $\mathfrak{sl}(3|2)$ as a sub-algebra, giving rise to the asymptotic symmetry, where the even\,-\,graded sector of the superalgebra decomposes into spin-2, the $\mathfrak{sl}(2)$ generators $\Lt_{i},(i=\pm1,0)$, a spin-1 element $\Jt$, a spin-2 multiplet $\At_{i},(i=\pm1,0)$ and a spin-3 multiplet $\Wt_{i},(i=\pm2,\pm1,0)$. All these generators together span the bosonic sub-algebra $\mathfrak{sl}(3)\oplus \mathfrak{sl}(2)\oplus \mathfrak{u}(1)$. Furthermore, the odd\,-\,graded elements decompose in two spin $\frac{3}{2}$ multiplets $\Gt_{r}^{M},(r=\pm\frac{1}{2}),\,(M=\pm)$ and two spin $\frac{5}{2}$ multiplets $\St_{r}^{M}, (r=\pm\frac{3}{2},\pm\frac{1}{2}),\,(M=\pm)$. Then, the bosonic sector of this algebra is given as follows:
\begin{empheq}{alignat=5}
	[\Lt_i,\Lt_j]&=(i-j)\Lt_{i+j}\,,
	 \qquad
	[\Lt_i,\At_j]=(i-j)\At_{i+j}\,,
    \\
    [\Lt_i,\Wt_j]&=(2i-j)\Wt_{i+j}\,,
     \qquad
	[\At_i,\At_j]=(i-j)\Lt_{i+j}\,,
	 \qquad
	[\At_i,\Wt_j]=(2i-j)\Wt_{i+j}\,,
	 \\
	[\Wt_i,\Wt_j]&=-\frac{1}{6}(i-j)(2i^2+2j^2-ij-8)(\Lt_{i+j}+A_{i+j})\,.
\end{empheq}
Additionally, the explicit commutation relations between the bosonic and fermionic sectors are given by
\begin{empheq}{alignat=5}
     \left[\Lt_{i},\Gt_{p}^{\pm}\right]&=\left(\frac{i}{2}-p\right)\Gt_{i+p}^{\pm},~~
	\qquad
    \left[\Lt_{i},\St_{p}^{\pm}\right]=\left(\frac{3i}{2}-p\right)\St_{i+p}^{\pm},~~
     \\
	[\Jt,\Gt_{r}^{\pm}]&=\pm\Gt_{r}^{\pm}\,,
	\qquad
	[\Jt,\St_{r}^{\pm}]=\pm\St_{r}^{\pm}\,,
	 \\
	[\At_i,\Gt_{r}^{\pm}]&=\frac{5}{3}\left(\frac{i}{2}-r\right)\Gt_{i+r}\pm\frac{4}{3}\St_{i+r}^{\pm}\,,
    \\
	[\At_i,\St_{r}^{\pm}]&=\frac{1}{3}\left(\frac{3i}{2}-r\right)\St_{i+r}^{\pm}\mp\frac{1}{3}\left(3i^2-2ir+r^2-\frac{9}{4}\right)\Gt_{i+r}^{\pm}\,,
 \end{empheq}
 \begin{empheq}{alignat=5}
	[\Wt_i,\Gt_r^{\pm}]&=-\frac{4}{3}\left(\frac{i}{2}-2r\right)\St_{i+r}^{\pm}\,,
		\\
	[\Wt_i,\St_r^{\pm}]&=\mp\frac{1}{3}\left(2r^2-2ri+i^2-\frac{5}{2}\right)\St_{i+r}^{\pm}\nonumber
	\\
	&-\frac{1}{6}\left(4r^3-3r^2i+2ri^2-i^3-9r+\frac{19}{4}i\right)\Gt_{i+r}^{\pm}.
	\end{empheq}
And lastly, the fermionic sector satisfy the following anti - commutation relations
\begin{empheq}{alignat=5}
	\left\{\Gt_{r}^{\pm},\Gt_{s}^{\mp}\right\} &=2\,\Lt_{r+s}\pm\left(r-s\right) \Jt
	\\
	\{\Gt_r^{\pm},\St_s^{\pm}\}&=-\frac{3}{2}\Wt_{r+s}^{\mp}+\frac{3}{4}(3r-s)\At_{r+s}-\frac{5}{4}(3r-s)\Lt_{r+s}\,,
	\\
	\{\St_r^{\pm},\St_s^{\mp}\}&=-\frac{3}{4}(r-s)\Wt_{r+s}+\frac{1}{8}\left(3s^2-4rs+3r^2-\frac{9}{2}\right)\left(L_{r+s}-3 \At_{r+s}\right)\nonumber
	\\
	&-\frac{1}{4}(r-s)\left(r^2+s^2-\frac{5}{2}\right)\Jt\,
\end{empheq}
except for zero commutators.

Having discussed the principle embedding of $\mathfrak{sl}(2|1)$ into $\mathfrak{sl}(3|2)$, we are now able to formulate the most general boundary conditions for asymptotically $AdS_3$ spacetimes. In accordance with this purpose it is useful to define the gauge connection as
\begin{eqnarray}\label{bouncondaf}
a_\varphi &=&
 \rho\mathcal{J}\Jt
+\gamma_i\mathcal{L}^i \Lt_{i}
+\vartheta_i\mathcal{A}^i \At_{i}
+\omega_i\mathcal{W}^i \Wt_{i}
+\sigma_M^p\mathcal{G}_{M}^{p}\Gt_{p}^{M}
+\tau_M^p\mathcal{S}_{M}^{p}\St_{p}^{M}\\
a_t &=&
\eta\Jt
+\mu^i \Lt_{i}
+\xi^i \At_{i}
+\mathit{f}^i \Wt_{i}
+\nu_M^p\Gt_{p}^{M}
+\psi_M^p\St_{p}^{M}
\end{eqnarray}
where
\begin{eqnarray}
\rho&=&\sigma_M^{\frac{1}{2}}=-\sigma_M^{\frac{1}{2}}=\frac{1}{k}\nonumber\\
2\gamma_1&=&2\gamma_{-1}-2\gamma_0=2\vartheta_1=2\vartheta_{-1}=-\vartheta_0=\frac{4}{k}\nonumber\\
3\tau_M^{-\frac{3}{2}}&=&-3\sigma_M^{\frac{3}{2}}=\tau_M^{\frac{1}{2}}=-\sigma_M^{\frac{1}{2}}=\frac{8}{3k}\nonumber\\
6\omega_{-2}=6\omega_{2}&=&-\frac{3}{2}\omega_{-1}=-\frac{3}{2}\omega_{1}=\omega_{0}=\frac{3}{k}\nonumber
\end{eqnarray}
are some scaling parameters. As a result, we have twenty four functions; twelve bosonic $\left(\mathcal{J},\mathcal{L}^i,\mathcal{A}^i,\mathcal{W}^i\right)$ and twelve fermionic $\left(\mathcal{G}_{M}^{p},\mathcal{S}_{M}^{p}\right)$ as $charges$. Also, we have in total twenty four independent functions $\left( \eta,\mu^i ,\xi^i  ,\mathit{f}^i  ,\nu_M^p ,\psi_M^p\right)$ too, as $chemical~~potentials$ for the time component.

In the presence of the loosest set of boundary conditions, thanks to the flatness conditions $(\ref{flatness})$, the equations of motion for fixed chemical potentials imposes the additional conditions as the temporal evolution of the twenty four independent source fields $\left( \mathcal{J} ,\mathcal{L}^i  ,\mathcal{A}^i  ,\mathcal{W}^i  ,\mathcal{G}_{M}^{p} ,\mathcal{S}_{M}^{p} \right)$ in the form
\begin{eqnarray}
\partial_{t}\mathcal{J}&=&
k\partial_{\varphi}\eta+\mathcal{G}_{2}^{\frac{1}{2}} \nu_{1}^{-\frac{1}{2}}+\mathcal{G}_{2}^{-\frac{1}{2}} \nu_{1}^{\frac{1}{2}}-\mathcal{G}_{1}^{\frac{1}{2}} \nu_{2}^{-\frac{1}{2}}-\mathcal{G}_{1}^{-\frac{1}{2}} \nu_{2}^{\frac{1}{2}}+\frac{4}{3} \mathcal{S}_{2}^{\frac{3}{2}} \psi_{1}^{-\frac{3}{2}}+\frac{4}{3} \mathcal{S}_{2}^{\frac{1}{2}} \psi_{1}^{-\frac{1}{2}}
+\frac{4}{3} \mathcal{S}_{2}^{-\frac{1}{2}} \psi_{1}^{\frac{1}{2}}+\frac{4}{3} \mathcal{S}_{2}^{-\frac{3}{2}} \psi_{1}^{\frac{3}{2}}\nonumber\\
&-&\frac{4}{3} \mathcal{S}_{1}^{\frac{3}{2}} \psi_{2}^{-\frac{3}{2}}-\frac{4}{3} \mathcal{S}_{1}^{\frac{1}{2}} \psi_{2}^{-\frac{1}{2}}-\frac{4}{3} \mathcal{S}_{1}^{-\frac{1}{2}} \psi_{2}^{\frac{1}{2}}-\frac{4}{3} \mathcal{S}_{1}^{-\frac{3}{2}} \psi_{2}^{\frac{3}{2}} \\
\partial_{t}\mathcal{L}^0&=&
\frac{k}{4}  \partial_{\varphi}\mu^0-\frac{1}{2} \mathcal{G}_{2}^{\frac{1}{2}} \nu_{1}^{-\frac{1}{2}}+\frac{1}{2} \mathcal{G}_{2}^{-\frac{1}{2}} \nu_{1}^{\frac{1}{2}}-\frac{1}{2} \mathcal{G}_{1}^{\frac{1}{2}} \nu_{2}^{-\frac{1}{2}}+\frac{1}{2} \mathcal{G}_{1}^{-\frac{1}{2}} \nu_{2}^{\frac{1}{2}}-\frac{5}{8} \mathcal{G}_{2}^{\frac{1}{2}} \psi_{1}^{-\frac{1}{2}}-\frac{5}{8} \mathcal{G}_{2}^{-\frac{1}{2}} \psi_{1}^{\frac{1}{2}}\nonumber\\
&+&\frac{5}{8} \mathcal{G}_{1}^{\frac{1}{2}} \psi_{2}^{-\frac{1}{2}}+\frac{5}{8} \mathcal{G}_{1}^{-\frac{1}{2}} \psi_{2}^{\frac{1}{2}}-\frac{5}{3} \nu_{1}^{-\frac{1}{2}} \mathcal{S}_{2}^{\frac{1}{2}}-\frac{5}{3} \nu_{1}^{\frac{1}{2}} \mathcal{S}_{2}^{-\frac{1}{2}}+\frac{5}{3} \nu_{2}^{-\frac{1}{2}} \mathcal{S}_{1}^{\frac{1}{2}}+\frac{5}{3} \nu_{2}^{\frac{1}{2}} \mathcal{S}_{1}^{-\frac{1}{2}}+\frac{1}{2} \mathcal{S}_{2}^{\frac{3}{2}} \psi_{1}^{-\frac{3}{2}}+\frac{1}{6} \mathcal{S}_{2}^{\frac{1}{2}} \psi_{1}^{-\frac{1}{2}}\nonumber\\
&-&\frac{1}{6} \mathcal{S}_{2}^{-\frac{1}{2}} \psi_{1}^{\frac{1}{2}}-\frac{1}{2} \mathcal{S}_{2}^{-\frac{3}{2}} \psi_{1}^{\frac{3}{2}}+\frac{1}{2} \mathcal{S}_{1}^{\frac{3}{2}} \psi_{2}^{-\frac{3}{2}}+\frac{1}{6} \mathcal{S}_{1}^{\frac{1}{2}} \psi_{2}^{-\frac{1}{2}}-\frac{1}{6} \mathcal{S}_{1}^{-\frac{1}{2}} \psi_{2}^{\frac{1}{2}}-\frac{1}{2} \mathcal{S}_{1}^{-\frac{3}{2}} \psi_{2}^{\frac{3}{2}}-\mathcal{A}^1 \xi^{-1}+\mathcal{A}^{-1} \xi^1\nonumber\\
&+&\mathit{f}^2 \mathcal{W}^{-2}+\frac{1}{2} \mathit{f}^1 \mathcal{W}^{-1}-\frac{1}{2} \mathit{f}^{-1} \mathcal{W}^1-\mathit{f}^{-2} \mathcal{W}^2-\mu^{-1} \mathcal{L}^1+\mu^1 \mathcal{L}^{-1}\\
\partial_{t}\mathcal{L}^{\pm1}&=&
-\frac{k}{2} \partial_{\varphi} \mu^{\pm1}\pm\mathcal{G}_{2}^{\pm\frac{1}{2}} \nu_{1}^{\pm\frac{1}{2}}\pm\mathcal{G}_{1}^{\pm\frac{1}{2}} \nu_{2}^{\pm\frac{1}{2}}+\frac{15}{8} \mathcal{G}_{2}^{\mp\frac{1}{2}} \psi_{1}^{\pm\frac{3}{2}}+\frac{5}{8} \mathcal{G}_{2}^{\pm\frac{1}{2}} \psi_{1}^{\pm\frac{1}{2}}-\frac{15}{8} \mathcal{G}_{\mp}^{\frac{1}{2}} \psi_{2}^{\pm\frac{3}{2}}-\frac{5}{8} \mathcal{G}_{1}^{\pm\frac{1}{2}} \psi_{2}^{\pm\frac{1}{2}}\nonumber\\
&-&\frac{5}{3} \nu_{1}^{\pm\frac{1}{2}} \mathcal{S}_{2}^{\pm\frac{1}{2}}
-\frac{5}{3} \nu_{1}^{\mp\frac{1}{2}} \mathcal{S}_{2}^{\pm\frac{3}{2}}+\frac{5}{3} \nu_{2}^{\pm\frac{1}{2}} \mathcal{S}_{1}^{\pm\frac{1}{2}}+\frac{5}{3} \nu_{2}^{\mp\frac{1}{2}} \mathcal{S}_{1}^{\pm\frac{3}{2}}\mp\mathcal{S}_{2}^{\mp\frac{1}{2}} \psi_{1}^{\pm\frac{3}{2}}\mp\frac{2}{3} \mathcal{S}_{2}^{\pm\frac{1}{2}} \psi_{1}^{\pm\frac{1}{2}}\mp\frac{1}{3} \mathcal{S}_{2}^{\pm\frac{3}{2}} \psi_{1}^{\mp\frac{1}{2}}\nonumber\\
&\mp&\mathcal{S}_{1}^{\mp\frac{1}{2}} \psi_{2}^{\pm\frac{3}{2}}
+\frac{2}{3} \mathcal{S}_{1}^{\pm\frac{1}{2}} \psi_{2}^{\pm\frac{1}{2}}
\mp\frac{1}{3} \mathcal{S}_{1}^{\pm\frac{3}{2}} \psi_{2}^{\mp\frac{1}{2}}\pm2 \mathcal{A}^0 \xi^{\pm1}\pm\mathcal{A}^{\pm1} \xi^0-\frac{1}{2} \mathit{f}^{\mp1} \mathcal{W}^{\pm2}\pm\mathit{f}^0 \mathcal{W}^{\pm1}\pm\frac{3}{2} \mathit{f}^{\pm1} \mathcal{W}^0\nonumber\\
&\pm&2 \mathit{f}^{\pm2} \mathcal{W}^{\mp1}\pm2 \mu^{\pm1} \mathcal{L}^0\pm\mu^0 \mathcal{L}^{\pm1}\\
\partial_{t}\mathcal{A}^0&=&
\frac{k}{4}\partial_{\varphi}\xi_0+\frac{3}{8} \mathcal{G}_{2}^{\frac{1}{2}} \psi _{1}^{-\frac{1}{2}}+\frac{3}{8} \mathcal{G}_{2}^{-\frac{1}{2}} \psi _{1}^{\frac{1}{2}}-\frac{3}{8} \mathcal{G}_{1}^{\frac{1}{2}} \psi _{2}^{-\frac{1}{2}}-\frac{3}{8} \mathcal{G}_{1}^{-\frac{1}{2}} \psi_{2}^{\frac{1}{2}}+\nu_{1}^{-\frac{1}{2}} \mathcal{S}_{2}^{\frac{1}{2}}+\nu _{1}^{\frac{1}{2}} \mathcal{S}_{2}^{-\frac{1}{2}}-\nu _{2}^{-\frac{1}{2}} \mathcal{S}_{1}^{\frac{1}{2}}\nonumber\\
&-&\nu _{2}^{\frac{1}{2}} \mathcal{S}_{1}^{-\frac{1}{2}}
-\frac{3}{2} \mathcal{S}_{2}^{\frac{3}{2}} \psi _{1}^{-\frac{3}{2}}-\frac{1}{2} \mathcal{S}_{2}^{\frac{1}{2}} \psi _{1}^{-\frac{1}{2}}+\frac{1}{2} \mathcal{S}_{2}^{-\frac{1}{2}} \psi _{1}^{\frac{1}{2}}+\frac{3}{2} \mathcal{S}_{2}^{-\frac{3}{2}} \psi _{1}^{\frac{3}{2}}-\frac{3}{2} \mathcal{S}_{1}^{\frac{3}{2}} \psi _{2}^{-\frac{3}{2}}-\frac{1}{2} \mathcal{S}_{1}^{\frac{1}{2}} \psi _{2}^{-\frac{1}{2}}\nonumber\\
&+&\frac{1}{2} \mathcal{S}_{1}^{-\frac{1}{2}} \psi _{2}^{\frac{1}{2}}
+\frac{3}{2} \mathcal{S}_{1}^{-\frac{3}{2}} \psi _{2}^{\frac{3}{2}}
-\mathcal{A}^1 \mu^{-1}+\mathcal{A}^{-1} \mu^1+\mathit{f}^2 \mathcal{W}^{-2}+\frac{1}{2} \mathit{f}^1 \mathcal{W}^{-1}-\frac{1}{2} \mathit{f}^{-1} \mathcal{W}^1-\mathit{f}^{-2} \mathcal{W}^2\nonumber\\
&-&\xi^{-1} \mathcal{L}^1+\xi^1 \mathcal{L}^{-1} \\
\partial_{t}\mathcal{A}^{\pm1}&=&
-\frac{k}{2}  \partial_{\varphi}\xi^{\pm1}-\frac{9}{8} \mathcal{G}_{2}^{\mp\frac{1}{2}} \psi_{1}^{\pm\frac{3}{2}}\pm\frac{3}{8} \mathcal{G}_{2}^{\pm\frac{1}{2}} \psi_{1}^{\pm\frac{1}{2}}+\frac{9}{8} \mathcal{G}_{1}^{\mp\frac{1}{2}} \psi_{2}^{\pm\frac{3}{2}}\mp\frac{3}{8} \mathcal{G}_{1}^{\pm\frac{1}{2}} \psi_{2}^{\pm\frac{1}{2}}+\nu_{1}^{\pm\frac{1}{2}} \mathcal{S}_{2}^{\pm\frac{1}{2}}+\nu_{1}^{\mp\frac{1}{2}} \mathcal{S}_{2}^{\pm\frac{3}{2}}\nonumber\\
&-&\nu_{2}^{\pm\frac{1}{2}} \mathcal{S}_{1}^{\pm\frac{1}{2}}
-\nu_{2}^{\mp\frac{1}{2}} \mathcal{S}_{1}^{\pm\frac{3}{2}}\pm3 \mathcal{S}_{2}^{\mp\frac{1}{2}} \psi_{1}^{\pm\frac{3}{2}}\pm2 \mathcal{S}_{2}^{\pm\frac{1}{2}} \psi_{1}^{\pm\frac{1}{2}}\pm\mathcal{S}_{2}^{\pm\frac{3}{2}} \psi_{1}^{\mp\frac{1}{2}}\pm3 \mathcal{S}_{1}^{\mp\frac{1}{2}} \psi_{2}^{\pm\frac{3}{2}}\pm2 \mathcal{S}_{1}^{\pm\frac{1}{2}} \psi_{2}^{\pm\frac{1}{2}}\nonumber\\
&\pm&\mathcal{S}_{1}^{\pm\frac{3}{2}} \psi_{2}^{\mp\frac{1}{2}}\pm2 \mathcal{A}^0 \mu^{\pm1}
\pm\mathcal{A}^{\pm1} \mu^0\pm\frac{1}{2} \mathit{f}^{\mp1} \mathcal{W}^{\pm2}\pm\mathit{f}^0 \mathcal{W}^{\pm1}\pm\frac{3}{2} \mathit{f}^{\pm1} \mathcal{W}^0\pm2 \mathit{f}^{\pm2} \mathcal{W}^{\mp1}
\pm 2\xi^{\pm1} \mathcal{L}^0\nonumber\\
&\pm&\xi^0 \mathcal{L}^{\pm1}\\
\partial_{t}\mathcal{W}^0&=&
\frac{k}{3} \partial_{\varphi}\mathit{f}^0 +\frac{1}{2} \mathcal{G}_{2}^{\frac{1}{2}} \psi_{1}^{-\frac{1}{2}}-\frac{1}{2} \mathcal{G}_{2}^{-\frac{1}{2}} \psi_{1}^{\frac{1}{2}}+\frac{1}{2} \mathcal{G}_{1}^{\frac{1}{2}} \psi_{2}^{-\frac{1}{2}}-\frac{1}{2} \mathcal{G}_{1}^{-\frac{1}{2}} \psi_{2}^{\frac{1}{2}}+\frac{4}{3} \nu_{1}^{-\frac{1}{2}} \mathcal{S}_{2}^{\frac{1}{2}}-\frac{4}{3} \nu_{1}^{\frac{1}{2}} \mathcal{S}_{2}^{-\frac{1}{2}}+\frac{4}{3} \nu_{2}^{-\frac{1}{2}} \mathcal{S}_{1}^{\frac{1}{2}}\nonumber\\
&-&\frac{4}{3} \nu_{2}^{\frac{1}{2}} \mathcal{S}_{1}^{-\frac{1}{2}}+\frac{2}{3} \mathcal{S}_{2}^{\frac{3}{2}} \psi_{1}^{-\frac{3}{2}}-\frac{2}{3} \mathcal{S}_{2}^{\frac{1}{2}} \psi_{1}^{-\frac{1}{2}}-\frac{2}{3} \mathcal{S}_{2}^{-\frac{1}{2}} \psi_{1}^{\frac{1}{2}}+\frac{2}{3} \mathcal{S}_{2}^{-\frac{3}{2}} \psi_{1}^{\frac{3}{2}}-\frac{2}{3} \mathcal{S}_{1}^{\frac{3}{2}} \psi_{2}^{-\frac{3}{2}}+\frac{2}{3} \mathcal{S}_{1}^{\frac{1}{2}} \psi_{2}^{-\frac{1}{2}}+\frac{2}{3} \mathcal{S}_{1}^{-\frac{1}{2}} \psi_{2}^{\frac{1}{2}}\nonumber\\
&-&\frac{2}{3} \mathcal{S}_{1}^{-\frac{3}{2}} \psi_{2}^{\frac{3}{2}}-2 \mathcal{A}^1 \mathit{f}^{-1}+2 \mathcal{A}^{-1} \mathit{f}^1-2 \mathit{f}^{-1} \mathcal{L}^1+2 \mathit{f}^1 \mathcal{L}^{-1} -2 \mu^{-1} \mathcal{W}^1+2 \mu^1 \mathcal{W}^{-1}-2 \xi^{-1} \mathcal{W}^1\nonumber\\
&+&2 \xi^1 \mathcal{W}^{-1}
\end{eqnarray}
\begin{eqnarray}
\partial_{t}\mathcal{W}^{\pm1}&=&
-\frac{k}{2} \partial_{\varphi} \mathit{f}^{\pm1}\pm\frac{3}{4} \mathcal{G}_{2}^{\mp\frac{1}{2}} \psi_{1}^{\pm\frac{3}{2}}\mp\frac{3}{4} \mathcal{G}_{2}^{\pm\frac{1}{2}} \psi_{1}^{\pm\frac{1}{2}}\pm\frac{3}{4} \mathcal{G}_{1}^{\mp\frac{1}{2}} \psi_{2}^{\pm\frac{3}{2}}\mp\frac{3}{4} \mathcal{G}_{1}^{\pm\frac{1}{2}} \psi_{2}^{\pm\frac{1}{2}}\mp2 \nu_{1}^{\pm\frac{1}{2}} \mathcal{S}_{2}^{\pm\frac{1}{2}}\pm\frac{2}{3} \nu_{1}^{\mp\frac{1}{2}} \mathcal{S}_{2}^{\pm\frac{3}{2}}\nonumber\\
&\mp&2 \nu_{2}^{\pm\frac{1}{2}} \mathcal{S}_{1}^{\pm\frac{1}{2}}\pm\frac{2}{3} \nu_{2}^{\mp\frac{1}{2}} \mathcal{S}_{1}^{\pm\frac{3}{2}}\mp2 \mathcal{S}_{2}^{\mp\frac{1}{2}} \psi_{1}^{\pm\frac{3}{2}}-\frac{2}{3} \mathcal{S}_{2}^{\pm\frac{3}{2}} \psi_{1}^{\mp\frac{1}{2}}\pm2 \mathcal{S}_{1}^{\mp\frac{1}{2}} \psi_{2}^{\pm\frac{3}{2}}+\frac{2}{3} \mathcal{S}_{1}^{\pm\frac{3}{2}} \psi_{2}^{\mp\frac{1}{2}}\mp4 \mathcal{A}^1 \mathit{f}^{\pm2}\nonumber\\
&\pm&2 \mathcal{A}^0 \mathit{f}^{\pm1}\pm2 \mathcal{A}^{\pm1} \mathit{f}^0\mp4 \mathit{f}^{\pm2} \mathcal{L}^{\mp1}\pm2 \mathit{f}^0 \mathcal{L}^{\pm1}\pm2 \mathit{f}^{\pm1} \mathcal{L}^0 \pm3 \mu^{\pm1} \mathcal{W}^0\pm\mu^0 \mathcal{W}^{\pm1}+\mu^{\mp1} \mathcal{W}^{\pm2}\nonumber\\
&\pm&3 \xi^{\pm1} \mathcal{W}^0\pm\xi^0 \mathcal{W}^{\pm1}
\mp\xi^{\mp1} \mathcal{W}^{\pm2}
\end{eqnarray}
\begin{eqnarray}
\partial_{t}\mathcal{W}^{\pm2}&=&
2 k\partial_{\varphi} \mathit{f}^{\pm2}\pm3 \mathcal{G}_{2}^{\pm\frac{1}{2}} \psi_{1}^{\pm\frac{3}{2}}\pm3 \mathcal{G}_{1}^{\pm\frac{1}{2}} \psi_{2}^{\pm\frac{3}{2}}\mp\frac{8}{3} \nu_{1}^{\pm\frac{1}{2}} \mathcal{S}_{2}^{\pm\frac{3}{2}}\mp\frac{8}{3} \nu_{2}^{\pm\frac{1}{2}} \mathcal{S}_{1}^{\pm\frac{3}{2}}+4 \mathcal{S}_{2}^{\pm\frac{1}{2}} \psi_{1}^{\pm\frac{3}{2}}\mp\frac{4}{3} \mathcal{S}_{2}^{\pm\frac{3}{2}} \psi_{1}^{\pm\frac{1}{2}}\nonumber\\
&-&4 \mathcal{S}_{1}^{\pm\frac{1}{2}} \psi_{2}^{\pm\frac{3}{2}}-\frac{4}{3} \mathcal{S}_{1}^{\pm\frac{3}{2}} \psi_{2}^{\pm\frac{1}{2}}\mp16 \mathcal{A}^0 \mathit{f}^{\pm2}\mp4 \mathcal{A}^{\pm1} \mathit{f}^{\pm1}+16 \mathit{f}^{\pm2} \mathcal{L}^0\mp4 \mathit{f}^{\pm1} \mathcal{L}^{\pm1}\pm4 \mu^{\pm1} \mathcal{W}^{\pm1}\nonumber\\
&\pm&2\mu^0 \mathcal{W}^{\pm2}\pm4 \xi^{\pm1} \mathcal{W}^{\pm1}\pm2 \xi^0 \mathcal{W}^{\pm2} \\
\partial_{t}\mathcal{G}_{M}^{\pm\frac{1}{2}}&=&\mp k\partial_{\varphi}\nu_{M}^{\pm\frac{1}{2}}+\frac{10}{3} \mathcal{A}^0 \nu_{M}^{\mp\frac{1}{2}}+\frac{10}{3} \mathcal{A}^{\pm1} \nu_{M}^{\mp\frac{1}{2}}
\mp4 \mathcal{A}^{\mp1} \psi_{M}^{\pm\frac{3}{2}}\mp\frac{8}{3} \mathcal{A}^0 \psi_{M}^{\pm\frac{1}{2}}\mp\frac{4}{3} \mathcal{A}^{\pm1} \psi_{M}^{\mp\frac{1}{2}}\mp\frac{8}{9} \mathit{f}^{\mp1} \mathcal{S}_{M}^{\pm\frac{3}{2}}\nonumber\\
&\mp&\frac{16}{9} \mathit{f}^0 \mathcal{S}_{M}^{\pm\frac{1}{2}}\mp\frac{8}{3} \mathit{f}^{\pm1} \mathcal{S}_{M}^{\mp\frac{1}{2}}\mp\frac{32}{9} \mathit{f}^{\pm2} \mathcal{S}_{M}^{\mp\frac{3}{2}}-\eta  \mathcal{G}_{M}^{\pm\frac{1}{2}}\pm\frac{1}{2} \mu^0 \mathcal{G}_{M}^{\pm\frac{1}{2}}\pm\mu^{\pm1} \mathcal{G}_{M}^{\mp\frac{1}{2}}\pm\frac{5}{6} \xi^0 \mathcal{G}_{M}^{\pm\frac{1}{2}}\pm\frac{5}{3} \xi^{\pm1} \mathcal{G}_{M}^{\mp\frac{1}{2}} \nonumber\\
&\mp&\mathcal{J} \nu_{M}^{\pm\frac{1}{2}}\mp\frac{16}{9} \xi^{\mp1} \mathcal{S}_{M}^{\pm\frac{3}{2}}-\frac{16}{9} \xi^0 \mathcal{S}_{M}^{\pm\frac{1}{2}}-\frac{16}{9} \xi^{\pm1} \mathcal{S}_{M}^{\mp\frac{1}{2}}-2
\mathcal{W}^{\mp1} \psi_{M}^{\pm\frac{3}{2}}-2 \mathcal{W}^0 \psi_{M}^{\pm\frac{1}{2}}-2 \mathcal{W}^{\pm1} \psi_{M}^{\mp\frac{1}{2}} \nonumber\\
&-& 2\mathcal{W}^{\pm2} \psi_{M}^{\mp\frac{3}{2}}+2 \mathcal{L}^0 \nu_{M}^{\pm\frac{1}{2}}+2 \mathcal{L}^{\pm1} \nu_{M}^{\mp\frac{1}{2}} \\
\partial_{t}\mathcal{S}_{M}^{\pm\frac{1}{2}}&=&\mp\frac{3}{8} k \psi_{M}^{\pm\frac{1}{2}}\mp2 \mathcal{A}^0 \nu_{M}^{\pm\frac{1}{2}}\pm\mathcal{A}^{\pm1} \nu_{M}^{\mp\frac{1}{2}}-\frac{3}{4} \mathcal{A}^{\mp1} \psi_{M}^{\pm\frac{3}{2}}+\frac{1}{4} \mathcal{A}^0 \psi_{M}^{\pm\frac{1}{2}}+\frac{1}{2} \mathcal{A}^{\pm1} \psi_{M}^{\mp\frac{1}{2}}\mp\frac{1}{2} \mathit{f}^0 \mathcal{G}_{M}^{\pm\frac{1}{2}}\nonumber\\
&\mp&\frac{3}{4} \mathit{f}^{\pm1} \mathcal{G}_{M}^{\mp\frac{1}{2} }
-\frac{2}{3} \mathit{f}^{\mp1} \mathcal{S}_{M}^{\pm\frac{3}{2}}-\frac{2}{3} \mathit{f}^0 \mathcal{S}_{M}^{\pm\frac{1}{2}}+\frac{4}{3} \mathit{f}^{\pm2} \mathcal{S}_{M}^{\mp\frac{3}{2}}-\frac{1}{2} \xi^0 \mathcal{G}_{M}^{\pm\frac{1}{2}}+\frac{1}{2} \xi^{\pm1} \mathcal{G}_{M}^{\mp\frac{1}{2}}\mp\frac{3}{8} \mathcal{J} \psi_{M}^{\pm\frac{1}{2}} -\eta  \mathcal{S}_{M}^{\pm\frac{1}{2}}\nonumber\\
&\pm&\mu^{\mp1} \mathcal{S}_{M}^{\pm\frac{3}{2}}\pm\frac{1}{2} \mu^0 \mathcal{S}_{M}^{\pm\frac{1}{2}}
-2 \mu^{\pm1} \mathcal{S}_{M}^{\mp\frac{1}{2}}\mp\frac{1}{3} \xi^{\mp1} \mathcal{S}_{M}^{\pm\frac{3}{2}}\pm\frac{1}{6} \xi^0 \mathcal{S}_{M}^{\pm\frac{1}{2}}\pm\frac{2}{3} \xi^{\pm1} \mathcal{S}_{M}^{\mp\frac{1}{2}}\pm\frac{3}{2} \mathcal{W}^0 \nu_{M}^{\pm\frac{1}{2}}-\frac{3}{2} \mathcal{W}^{\pm1} \nu_{M}^{\mp\frac{1}{2}}\nonumber\\
&+&\frac{3}{2} \mathcal{W}^{\mp1} \psi_{M}^{\pm\frac{3}{2}}\mp\frac{3}{4} \mathcal{W}^0 \psi_{M}^{\pm\frac{1}{2}}
\pm\frac{3}{4} \mathcal{W}^{\pm2} \psi_{M}^{\mp\frac{3}{2}}-\frac{9}{4} \mathcal{L}^{\mp1} \psi_{M}^{\pm\frac{3}{2}}+\frac{3}{4} \mathcal{L}^0 \psi_{M}^{\pm\frac{1}{2}}+\frac{3}{2} \mathcal{L}^{\pm1} \psi_{M}^{\mp\frac{1}{2}}\\
\partial_{t}\mathcal{S}_{M}^{\pm\frac{3}{2}}&=&
\pm\frac{9}{8} k \psi_{M}^{\pm\frac{3}{2}}
\mp3 \mathcal{A}^{-1} \nu_{M}^{\pm\frac{1}{2}}
-\frac{9}{4} \mathcal{A}^0 \psi_{M}^{\pm\frac{3}{2}}
-\frac{3}{4} \mathcal{A}^{\pm1} \psi_{M}^{\pm\frac{1}{2}}
\pm\frac{3}{4} \mathit{f}^{\pm1} \mathcal{G}_{M}^{\pm\frac{1}{2}}
\pm3 \mathit{f}^{\pm2} \mathcal{G}_{M}^{\mp\frac{1}{2}}
+\frac{2}{3} \mathit{f}^0 \mathcal{S}_{M}^{\pm\frac{3}{2}}\nonumber\\
&+&2 \mathit{f}^{\pm1} \mathcal{S}_{M}^{\pm\frac{1}{2}}
+4 \mathit{f}^{\pm2} \mathcal{S}_{M}^{\mp\frac{1}{2}}
+\frac{3}{2} \xi^{\pm1} \mathcal{G}_{M}^{\pm\frac{1}{2}}
\pm\frac{9}{8} \mathcal{J} \psi_{M}^{\pm\frac{3}{2}}
-\eta  \mathcal{S}_{M}^{\pm\frac{3}{2}}
\pm\frac{3}{2} \mu^0 \mathcal{S}_{M}^{\pm\frac{3}{2}}
\pm3 \mu^{\pm1} \mathcal{S}_{M}^{\pm\frac{1}{2}}
\pm\frac{1}{2} \xi^0 \mathcal{S}_{M}^{\pm\frac{3}{2}}\nonumber\\
&\pm&\xi^{\pm1} \mathcal{S}_{M}^{\pm\frac{1}{2}}
-\frac{3}{2} \mathcal{W}^{\pm1} \nu_{M}^{\pm\frac{1}{2}}
-\frac{3}{2} \mathcal{W}^{\pm2} \nu_{M}^{\mp\frac{1}{2}}
+\frac{9}{4} \mathcal{W}^0 \psi_{M}^{\pm\frac{3}{2}}
\mp\frac{3}{2} \mathcal{W}^{\pm1} \psi_{M}^{\pm\frac{1}{2}}
\mp\frac{3}{4} \mathcal{W}^{\pm2} \psi_{M}^{\mp\frac{1}{2}}\nonumber\\
&-&\frac{27}{4} \mathcal{L}^0 \psi_{M}^{\pm\frac{3}{2}}
-\frac{9}{4} \mathcal{L}^{\pm1} \psi_{M}^{\pm\frac{1}{2}}
\end{eqnarray}
Thus, the consequences of our calculations to derive the relevant superalgebra for the loosest set of boundary conditions can now be evaluated through a canonical analysis. We now consider the boundary preserving gauge transformations (encompassing all) $(\ref{boundarycond})$ generated by the  $\mathfrak{sl}(3|2)$ Lie superalgebra-valued gauge parameter $\lambda$, which we choose as
\begin{equation}
 \lambda=b^{-1}
                     \bigg[
\varrho\Jt
+\epsilon^i \Lt_{i}
+\phi^i \At_{i}
+\upsilon^i \Wt_{i}
+\varsigma_M^p\Gt_{p}^{M}
+\omega_M^p\St_{p}^{M}
                     \bigg]b.\label{boundarycond2221}
\end{equation}
Note that there are in total twenty four arbitrary  functions on the boundary, consist of twelve bosonic $(\varrho,\,\mathcal{\epsilon}^i,\,{\phi}^i,\,{\upsilon}^i)$ and twelve fermionic $(\mathcal{\varsigma}_{M}^{p},\,\mathcal{\omega}_{M}^{p})$. Inserting this expression into $(\ref{boundarycond})$ imposes the gauge transformations as
\begin{eqnarray}
\partial_{\lambda}\mathcal{J}&=&
k\partial_{\varphi}\varrho+\mathcal{G}_{2}^{\frac{1}{2}} \nu_{1}^{-\frac{1}{2}}+\mathcal{G}_{2}^{-\frac{1}{2}} \nu_{1}^{\frac{1}{2}}-\mathcal{G}_{1}^{\frac{1}{2}} \nu_{2}^{-\frac{1}{2}}-\mathcal{G}_{1}^{-\frac{1}{2}} \nu_{2}^{\frac{1}{2}}+\frac{4}{3} \mathcal{S}_{2}^{\frac{3}{2}} \omega_{1}^{-\frac{3}{2}}+\frac{4}{3} \mathcal{S}_{2}^{\frac{1}{2}} \omega_{1}^{-\frac{1}{2}}
+\frac{4}{3} \mathcal{S}_{2}^{-\frac{1}{2}} \omega_{1}^{\frac{1}{2}}+\frac{4}{3} \mathcal{S}_{2}^{-\frac{3}{2}} \omega_{1}^{\frac{3}{2}}\nonumber\\
&-&\frac{4}{3} \mathcal{S}_{1}^{\frac{3}{2}} \omega_{2}^{-\frac{3}{2}}-\frac{4}{3} \mathcal{S}_{1}^{\frac{1}{2}} \omega_{2}^{-\frac{1}{2}}-\frac{4}{3} \mathcal{S}_{1}^{-\frac{1}{2}} \omega_{2}^{\frac{1}{2}}-\frac{4}{3} \mathcal{S}_{1}^{-\frac{3}{2}} \omega_{2}^{\frac{3}{2}},\label{traqq1} \\
\partial_{\lambda}\mathcal{L}^0&=&
\frac{k}{4}  \partial_{\varphi}\epsilon^0-\frac{1}{2} \mathcal{G}_{2}^{\frac{1}{2}} \nu_{1}^{-\frac{1}{2}}+\frac{1}{2} \mathcal{G}_{2}^{-\frac{1}{2}} \nu_{1}^{\frac{1}{2}}-\frac{1}{2} \mathcal{G}_{1}^{\frac{1}{2}} \nu_{2}^{-\frac{1}{2}}+\frac{1}{2} \mathcal{G}_{1}^{-\frac{1}{2}} \nu_{2}^{\frac{1}{2}}-\frac{5}{8} \mathcal{G}_{2}^{\frac{1}{2}} \omega_{1}^{-\frac{1}{2}}-\frac{5}{8} \mathcal{G}_{2}^{-\frac{1}{2}} \omega_{1}^{\frac{1}{2}}\nonumber\\
&+&\frac{5}{8} \mathcal{G}_{1}^{\frac{1}{2}} \omega_{2}^{-\frac{1}{2}}+\frac{5}{8} \mathcal{G}_{1}^{-\frac{1}{2}} \omega_{2}^{\frac{1}{2}}-\frac{5}{3} \nu_{1}^{-\frac{1}{2}} \mathcal{S}_{2}^{\frac{1}{2}}-\frac{5}{3} \nu_{1}^{\frac{1}{2}} \mathcal{S}_{2}^{-\frac{1}{2}}+\frac{5}{3} \nu_{2}^{-\frac{1}{2}} \mathcal{S}_{1}^{\frac{1}{2}}+\frac{5}{3} \nu_{2}^{\frac{1}{2}} \mathcal{S}_{1}^{-\frac{1}{2}}+\frac{1}{2} \mathcal{S}_{2}^{\frac{3}{2}} \omega_{1}^{-\frac{3}{2}}+\frac{1}{6} \mathcal{S}_{2}^{\frac{1}{2}} \omega_{1}^{-\frac{1}{2}}\nonumber\\
&-&\frac{1}{6} \mathcal{S}_{2}^{-\frac{1}{2}} \omega_{1}^{\frac{1}{2}}-\frac{1}{2} \mathcal{S}_{2}^{-\frac{3}{2}} \omega_{1}^{\frac{3}{2}}+\frac{1}{2} \mathcal{S}_{1}^{\frac{3}{2}} \omega_{2}^{-\frac{3}{2}}+\frac{1}{6} \mathcal{S}_{1}^{\frac{1}{2}} \omega_{2}^{-\frac{1}{2}}-\frac{1}{6} \mathcal{S}_{1}^{-\frac{1}{2}} \omega_{2}^{\frac{1}{2}}-\frac{1}{2} \mathcal{S}_{1}^{-\frac{3}{2}} \omega_{2}^{\frac{3}{2}}-\mathcal{A}^1 \phi^{-1}+\mathcal{A}^{-1} \phi^1\nonumber\\
&+&\upsilon^2 \mathcal{W}^{-2}+\frac{1}{2} \upsilon^1 \mathcal{W}^{-1}-\frac{1}{2} \upsilon^{-1} \mathcal{W}^1-\upsilon^{-2} \mathcal{W}^2-\epsilon^{-1} \mathcal{L}^1+\epsilon^1 \mathcal{L}^{-1} \\
\partial_{\lambda}\mathcal{L}^{\pm1}&=&
-\frac{k}{2} \partial_{\varphi} \epsilon^{\pm1}\pm\mathcal{G}_{2}^{\pm\frac{1}{2}} \nu_{1}^{\pm\frac{1}{2}}\pm\mathcal{G}_{1}^{\pm\frac{1}{2}} \nu_{2}^{\pm\frac{1}{2}}+\frac{15}{8} \mathcal{G}_{2}^{\mp\frac{1}{2}} \omega_{1}^{\pm\frac{3}{2}}+\frac{5}{8} \mathcal{G}_{2}^{\pm\frac{1}{2}} \omega_{1}^{\pm\frac{1}{2}}-\frac{15}{8} \mathcal{G}_{\mp}^{\frac{1}{2}} \omega_{2}^{\pm\frac{3}{2}}-\frac{5}{8} \mathcal{G}_{1}^{\pm\frac{1}{2}} \omega_{2}^{\pm\frac{1}{2}}\nonumber\\
&-&\frac{5}{3} \nu_{1}^{\pm\frac{1}{2}} \mathcal{S}_{2}^{\pm\frac{1}{2}}
-\frac{5}{3} \nu_{1}^{\mp\frac{1}{2}} \mathcal{S}_{2}^{\pm\frac{3}{2}}+\frac{5}{3} \nu_{2}^{\pm\frac{1}{2}} \mathcal{S}_{1}^{\pm\frac{1}{2}}+\frac{5}{3} \nu_{2}^{\mp\frac{1}{2}} \mathcal{S}_{1}^{\pm\frac{3}{2}}\mp\mathcal{S}_{2}^{\mp\frac{1}{2}} \omega_{1}^{\pm\frac{3}{2}}\mp\frac{2}{3} \mathcal{S}_{2}^{\pm\frac{1}{2}} \omega_{1}^{\pm\frac{1}{2}}\mp\frac{1}{3} \mathcal{S}_{2}^{\pm\frac{3}{2}} \omega_{1}^{\mp\frac{1}{2}}\nonumber\\
&\mp&\mathcal{S}_{1}^{\mp\frac{1}{2}} \omega_{2}^{\pm\frac{3}{2}}
+\frac{2}{3} \mathcal{S}_{1}^{\pm\frac{1}{2}} \omega_{2}^{\pm\frac{1}{2}}
\mp\frac{1}{3} \mathcal{S}_{1}^{\pm\frac{3}{2}} \omega_{2}^{\mp\frac{1}{2}}\pm2 \mathcal{A}^0 \phi^{\pm1}\pm\mathcal{A}^{\pm1} \phi^0-\frac{1}{2} \upsilon^{\mp1} \mathcal{W}^{\pm2}\pm\upsilon^0 \mathcal{W}^{\pm1}\pm\frac{3}{2} \upsilon^{\pm1} \mathcal{W}^0\nonumber\\
&\pm&2 \upsilon^{\pm2} \mathcal{W}^{\mp1}\pm2 \epsilon^{\pm1} \mathcal{L}^0\pm\epsilon^0 \mathcal{L}^{\pm1}\\
\partial_{\lambda}\mathcal{A}^0&=&
\frac{k}{4}\partial_{\varphi}\phi_0+\frac{3}{8} \mathcal{G}_{2}^{\frac{1}{2}} \omega _{1}^{-\frac{1}{2}}+\frac{3}{8} \mathcal{G}_{2}^{-\frac{1}{2}} \omega _{1}^{\frac{1}{2}}-\frac{3}{8} \mathcal{G}_{1}^{\frac{1}{2}} \omega _{2}^{-\frac{1}{2}}-\frac{3}{8} \mathcal{G}_{1}^{-\frac{1}{2}} \omega_{2}^{\frac{1}{2}}+\nu_{1}^{-\frac{1}{2}} \mathcal{S}_{2}^{\frac{1}{2}}+\nu _{1}^{\frac{1}{2}} \mathcal{S}_{2}^{-\frac{1}{2}}-\nu _{2}^{-\frac{1}{2}} \mathcal{S}_{1}^{\frac{1}{2}}\nonumber\\
&-&\nu _{2}^{\frac{1}{2}} \mathcal{S}_{1}^{-\frac{1}{2}}
-\frac{3}{2} \mathcal{S}_{2}^{\frac{3}{2}} \omega _{1}^{-\frac{3}{2}}-\frac{1}{2} \mathcal{S}_{2}^{\frac{1}{2}} \omega _{1}^{-\frac{1}{2}}+\frac{1}{2} \mathcal{S}_{2}^{-\frac{1}{2}} \omega _{1}^{\frac{1}{2}}+\frac{3}{2} \mathcal{S}_{2}^{-\frac{3}{2}} \omega _{1}^{\frac{3}{2}}-\frac{3}{2} \mathcal{S}_{1}^{\frac{3}{2}} \omega _{2}^{-\frac{3}{2}}-\frac{1}{2} \mathcal{S}_{1}^{\frac{1}{2}} \omega _{2}^{-\frac{1}{2}}\nonumber\\
&+&\frac{1}{2} \mathcal{S}_{1}^{-\frac{1}{2}} \omega _{2}^{\frac{1}{2}}
+\frac{3}{2} \mathcal{S}_{1}^{-\frac{3}{2}} \omega _{2}^{\frac{3}{2}}
-\mathcal{A}^1 \epsilon^{-1}+\mathcal{A}^{-1} \epsilon^1+\upsilon^2 \mathcal{W}^{-2}+\frac{1}{2} \upsilon^1 \mathcal{W}^{-1}-\frac{1}{2} \upsilon^{-1} \mathcal{W}^1-\upsilon^{-2} \mathcal{W}^2\nonumber\\
&-&\phi^{-1} \mathcal{L}^1+\phi^1 \mathcal{L}^{-1} \\
\partial_{\lambda}\mathcal{A}^{\pm1}&=&
-\frac{k}{2}  \partial_{\varphi}\phi^{\pm1}-\frac{9}{8} \mathcal{G}_{2}^{\mp\frac{1}{2}} \omega_{1}^{\pm\frac{3}{2}}\pm\frac{3}{8} \mathcal{G}_{2}^{\pm\frac{1}{2}} \omega_{1}^{\pm\frac{1}{2}}+\frac{9}{8} \mathcal{G}_{1}^{\mp\frac{1}{2}} \omega_{2}^{\pm\frac{3}{2}}\mp\frac{3}{8} \mathcal{G}_{1}^{\pm\frac{1}{2}} \omega_{2}^{\pm\frac{1}{2}}+\nu_{1}^{\pm\frac{1}{2}} \mathcal{S}_{2}^{\pm\frac{1}{2}}+\nu_{1}^{\mp\frac{1}{2}} \mathcal{S}_{2}^{\pm\frac{3}{2}}\nonumber\\
&-&\nu_{2}^{\pm\frac{1}{2}} \mathcal{S}_{1}^{\pm\frac{1}{2}}
-\nu_{2}^{\mp\frac{1}{2}} \mathcal{S}_{1}^{\pm\frac{3}{2}}\pm3 \mathcal{S}_{2}^{\mp\frac{1}{2}} \omega_{1}^{\pm\frac{3}{2}}\pm2 \mathcal{S}_{2}^{\pm\frac{1}{2}} \omega_{1}^{\pm\frac{1}{2}}\pm\mathcal{S}_{2}^{\pm\frac{3}{2}} \omega_{1}^{\mp\frac{1}{2}}\pm3 \mathcal{S}_{1}^{\mp\frac{1}{2}} \omega_{2}^{\pm\frac{3}{2}}\pm2 \mathcal{S}_{1}^{\pm\frac{1}{2}} \omega_{2}^{\pm\frac{1}{2}}\nonumber\\
&\pm&\mathcal{S}_{1}^{\pm\frac{3}{2}} \omega_{2}^{\mp\frac{1}{2}}\pm2 \mathcal{A}^0 \epsilon^{\pm1}
\pm\mathcal{A}^{\pm1} \epsilon^0\pm\frac{1}{2} \upsilon^{\mp1} \mathcal{W}^{\pm2}\pm\upsilon^0 \mathcal{W}^{\pm1}\pm\frac{3}{2} \upsilon^{\pm1} \mathcal{W}^0\pm2 \upsilon^{\pm2} \mathcal{W}^{\mp1}
\pm 2\phi^{\pm1} \mathcal{L}^0\nonumber\\
&\pm&\phi^0 \mathcal{L}^{\pm1}\\
\partial_{\lambda}\mathcal{W}^0&=&
\frac{k}{3} \partial_{\varphi}\upsilon^0 +\frac{1}{2} \mathcal{G}_{2}^{\frac{1}{2}} \omega_{1}^{-\frac{1}{2}}-\frac{1}{2} \mathcal{G}_{2}^{-\frac{1}{2}} \omega_{1}^{\frac{1}{2}}+\frac{1}{2} \mathcal{G}_{1}^{\frac{1}{2}} \omega_{2}^{-\frac{1}{2}}-\frac{1}{2} \mathcal{G}_{1}^{-\frac{1}{2}} \omega_{2}^{\frac{1}{2}}+\frac{4}{3} \nu_{1}^{-\frac{1}{2}} \mathcal{S}_{2}^{\frac{1}{2}}-\frac{4}{3} \nu_{1}^{\frac{1}{2}} \mathcal{S}_{2}^{-\frac{1}{2}}+\frac{4}{3} \nu_{2}^{-\frac{1}{2}} \mathcal{S}_{1}^{\frac{1}{2}}\nonumber\\
&-&\frac{4}{3} \nu_{2}^{\frac{1}{2}} \mathcal{S}_{1}^{-\frac{1}{2}}+\frac{2}{3} \mathcal{S}_{2}^{\frac{3}{2}} \omega_{1}^{-\frac{3}{2}}-\frac{2}{3} \mathcal{S}_{2}^{\frac{1}{2}} \omega_{1}^{-\frac{1}{2}}-\frac{2}{3} \mathcal{S}_{2}^{-\frac{1}{2}} \omega_{1}^{\frac{1}{2}}+\frac{2}{3} \mathcal{S}_{2}^{-\frac{3}{2}} \omega_{1}^{\frac{3}{2}}-\frac{2}{3} \mathcal{S}_{1}^{\frac{3}{2}} \omega_{2}^{-\frac{3}{2}}+\frac{2}{3} \mathcal{S}_{1}^{\frac{1}{2}} \omega_{2}^{-\frac{1}{2}}+\frac{2}{3} \mathcal{S}_{1}^{-\frac{1}{2}} \omega_{2}^{\frac{1}{2}}\nonumber\\
&-&\frac{2}{3} \mathcal{S}_{1}^{-\frac{3}{2}} \omega_{2}^{\frac{3}{2}}-2 \mathcal{A}^1 \upsilon^{-1}+2 \mathcal{A}^{-1} \upsilon^1-2 \upsilon^{-1} \mathcal{L}^1+2 \upsilon^1 \mathcal{L}^{-1} -2 \epsilon^{-1} \mathcal{W}^1+2 \epsilon^1 \mathcal{W}^{-1}-2 \phi^{-1} \mathcal{W}^1\nonumber\\
&+&2 \phi^1 \mathcal{W}^{-1}
\end{eqnarray}
\begin{eqnarray}
\partial_{\lambda}\mathcal{W}^{\pm1}&=&
-\frac{k}{2} \partial_{\varphi} \upsilon^{\pm1}\pm\frac{3}{4} \mathcal{G}_{2}^{\mp\frac{1}{2}} \omega_{1}^{\pm\frac{3}{2}}\mp\frac{3}{4} \mathcal{G}_{2}^{\pm\frac{1}{2}} \omega_{1}^{\pm\frac{1}{2}}\pm\frac{3}{4} \mathcal{G}_{1}^{\mp\frac{1}{2}} \omega_{2}^{\pm\frac{3}{2}}\mp\frac{3}{4} \mathcal{G}_{1}^{\pm\frac{1}{2}} \omega_{2}^{\pm\frac{1}{2}}\mp2 \nu_{1}^{\pm\frac{1}{2}} \mathcal{S}_{2}^{\pm\frac{1}{2}}\pm\frac{2}{3} \nu_{1}^{\mp\frac{1}{2}} \mathcal{S}_{2}^{\pm\frac{3}{2}}\nonumber\\
&\mp&2 \nu_{2}^{\pm\frac{1}{2}} \mathcal{S}_{1}^{\pm\frac{1}{2}}\pm\frac{2}{3} \nu_{2}^{\mp\frac{1}{2}} \mathcal{S}_{1}^{\pm\frac{3}{2}}\mp2 \mathcal{S}_{2}^{\mp\frac{1}{2}} \omega_{1}^{\pm\frac{3}{2}}-\frac{2}{3} \mathcal{S}_{2}^{\pm\frac{3}{2}} \omega_{1}^{\mp\frac{1}{2}}\pm2 \mathcal{S}_{1}^{\mp\frac{1}{2}} \omega_{2}^{\pm\frac{3}{2}}+\frac{2}{3} \mathcal{S}_{1}^{\pm\frac{3}{2}} \omega_{2}^{\mp\frac{1}{2}}\mp4 \mathcal{A}^1 \upsilon^{\pm2}\nonumber\\
&\pm&2 \mathcal{A}^0 \upsilon^{\pm1}\pm2 \mathcal{A}^{\pm1} \upsilon^0\mp4 \upsilon^{\pm2} \mathcal{L}^{\mp1}\pm2 \upsilon^0 \mathcal{L}^{\pm1}\pm2 \upsilon^{\pm1} \mathcal{L}^0 \pm3 \epsilon^{\pm1} \mathcal{W}^0\pm\epsilon^0 \mathcal{W}^{\pm1}+\epsilon^{\mp1} \mathcal{W}^{\pm2}\nonumber\\
&\pm&3 \phi^{\pm1} \mathcal{W}^0\pm\phi^0 \mathcal{W}^{\pm1}
\mp\phi^{\mp1} \mathcal{W}^{\pm2}
\end{eqnarray}
\begin{eqnarray}
\partial_{\lambda}\mathcal{W}^{\pm2}&=&
2 k\partial_{\varphi} \upsilon^{\pm2}\pm3 \mathcal{G}_{2}^{\pm\frac{1}{2}} \omega_{1}^{\pm\frac{3}{2}}\pm3 \mathcal{G}_{1}^{\pm\frac{1}{2}} \omega_{2}^{\pm\frac{3}{2}}\mp\frac{8}{3} \nu_{1}^{\pm\frac{1}{2}} \mathcal{S}_{2}^{\pm\frac{3}{2}}\mp\frac{8}{3} \nu_{2}^{\pm\frac{1}{2}} \mathcal{S}_{1}^{\pm\frac{3}{2}}+4 \mathcal{S}_{2}^{\pm\frac{1}{2}} \omega_{1}^{\pm\frac{3}{2}}\mp\frac{4}{3} \mathcal{S}_{2}^{\pm\frac{3}{2}} \omega_{1}^{\pm\frac{1}{2}}\nonumber\\
&-&4 \mathcal{S}_{1}^{\pm\frac{1}{2}} \omega_{2}^{\pm\frac{3}{2}}-\frac{4}{3} \mathcal{S}_{1}^{\pm\frac{3}{2}} \omega_{2}^{\pm\frac{1}{2}}\mp16 \mathcal{A}^0 \upsilon^{\pm2}\mp4 \mathcal{A}^{\pm1} \upsilon^{\pm1}+16 \upsilon^{\pm2} \mathcal{L}^0\mp4 \upsilon^{\pm1} \mathcal{L}^{\pm1}\pm4 \epsilon^{\pm1} \mathcal{W}^{\pm1}\nonumber\\
&\pm&2\epsilon^0 \mathcal{W}^{\pm2}\pm4 \phi^{\pm1} \mathcal{W}^{\pm1}\pm2 \phi^0 \mathcal{W}^{\pm2} \\
\partial_{\lambda}\mathcal{G}_{M}^{\pm\frac{1}{2}}&=&\mp k\partial_{\varphi}\nu_{M}^{\pm\frac{1}{2}}+\frac{10}{3} \mathcal{A}^0 \nu_{M}^{\mp\frac{1}{2}}+\frac{10}{3} \mathcal{A}^{\pm1} \nu_{M}^{\mp\frac{1}{2}}
\mp4 \mathcal{A}^{\mp1} \omega_{M}^{\pm\frac{3}{2}}\mp\frac{8}{3} \mathcal{A}^0 \omega_{M}^{\pm\frac{1}{2}}\mp\frac{4}{3} \mathcal{A}^{\pm1} \omega_{M}^{\mp\frac{1}{2}}\mp\frac{8}{9} \upsilon^{\mp1} \mathcal{S}_{M}^{\pm\frac{3}{2}}\nonumber\\
&\mp&\frac{16}{9} \upsilon^0 \mathcal{S}_{M}^{\pm\frac{1}{2}}\mp\frac{8}{3} \upsilon^{\pm1} \mathcal{S}_{M}^{\mp\frac{1}{2}}\mp\frac{32}{9} \upsilon^{\pm2} \mathcal{S}_{M}^{\mp\frac{3}{2}}-\varrho  \mathcal{G}_{M}^{\pm\frac{1}{2}}\pm\frac{1}{2} \epsilon^0 \mathcal{G}_{M}^{\pm\frac{1}{2}}\pm\epsilon^{\pm1} \mathcal{G}_{M}^{\mp\frac{1}{2}}\pm\frac{5}{6} \phi^0 \mathcal{G}_{M}^{\pm\frac{1}{2}}\pm\frac{5}{3} \phi^{\pm1} \mathcal{G}_{M}^{\mp\frac{1}{2}} \nonumber\\
&\mp&\mathcal{J} \nu_{M}^{\pm\frac{1}{2}}\mp\frac{16}{9} \phi^{\mp1} \mathcal{S}_{M}^{\pm\frac{3}{2}}-\frac{16}{9} \phi^0 \mathcal{S}_{M}^{\pm\frac{1}{2}}-\frac{16}{9} \phi^{\pm1} \mathcal{S}_{M}^{\mp\frac{1}{2}}-2
\mathcal{W}^{\mp1} \omega_{M}^{\pm\frac{3}{2}}-2 \mathcal{W}^0 \omega_{M}^{\pm\frac{1}{2}}-2 \mathcal{W}^{\pm1} \omega_{M}^{\mp\frac{1}{2}} \nonumber\\
&-& 2\mathcal{W}^{\pm2} \omega_{M}^{\mp\frac{3}{2}}+2 \mathcal{L}^0 \nu_{M}^{\pm\frac{1}{2}}+2 \mathcal{L}^{\pm1} \nu_{M}^{\mp\frac{1}{2}} \\
\partial_{\lambda}\mathcal{S}_{M}^{\pm\frac{1}{2}}&=&\mp\frac{3}{8} k \omega_{M}^{\pm\frac{1}{2}}\mp2 \mathcal{A}^0 \nu_{M}^{\pm\frac{1}{2}}\pm\mathcal{A}^{\pm1} \nu_{M}^{\mp\frac{1}{2}}-\frac{3}{4} \mathcal{A}^{\mp1} \omega_{M}^{\pm\frac{3}{2}}+\frac{1}{4} \mathcal{A}^0 \omega_{M}^{\pm\frac{1}{2}}+\frac{1}{2} \mathcal{A}^{\pm1} \omega_{M}^{\mp\frac{1}{2}}\mp\frac{1}{2} \upsilon^0 \mathcal{G}_{M}^{\pm\frac{1}{2}}\nonumber\\
&\mp&\frac{3}{4} \upsilon^{\pm1} \mathcal{G}_{M}^{\mp\frac{1}{2} }
-\frac{2}{3} \upsilon^{\mp1} \mathcal{S}_{M}^{\pm\frac{3}{2}}-\frac{2}{3} \upsilon^0 \mathcal{S}_{M}^{\pm\frac{1}{2}}+\frac{4}{3} \upsilon^{\pm2} \mathcal{S}_{M}^{\mp\frac{3}{2}}-\frac{1}{2} \phi^0 \mathcal{G}_{M}^{\pm\frac{1}{2}}+\frac{1}{2} \phi^{\pm1} \mathcal{G}_{M}^{\mp\frac{1}{2}}\mp\frac{3}{8} \mathcal{J} \omega_{M}^{\pm\frac{1}{2}} -\varrho  \mathcal{S}_{M}^{\pm\frac{1}{2}}\nonumber\\
&\pm&\epsilon^{\mp1} \mathcal{S}_{M}^{\pm\frac{3}{2}}\pm\frac{1}{2} \epsilon^0 \mathcal{S}_{M}^{\pm\frac{1}{2}}
-2 \epsilon^{\pm1} \mathcal{S}_{M}^{\mp\frac{1}{2}}\mp\frac{1}{3} \phi^{\mp1} \mathcal{S}_{M}^{\pm\frac{3}{2}}\pm\frac{1}{6} \phi^0 \mathcal{S}_{M}^{\pm\frac{1}{2}}\pm\frac{2}{3} \phi^{\pm1} \mathcal{S}_{M}^{\mp\frac{1}{2}}\pm\frac{3}{2} \mathcal{W}^0 \nu_{M}^{\pm\frac{1}{2}}-\frac{3}{2} \mathcal{W}^{\pm1} \nu_{M}^{\mp\frac{1}{2}}\nonumber\\
&+&\frac{3}{2} \mathcal{W}^{\mp1} \omega_{M}^{\pm\frac{3}{2}}\mp\frac{3}{4} \mathcal{W}^0 \omega_{M}^{\pm\frac{1}{2}}
\pm\frac{3}{4} \mathcal{W}^{\pm2} \omega_{M}^{\mp\frac{3}{2}}-\frac{9}{4} \mathcal{L}^{\mp1} \omega_{M}^{\pm\frac{3}{2}}+\frac{3}{4} \mathcal{L}^0 \omega_{M}^{\pm\frac{1}{2}}+\frac{3}{2} \mathcal{L}^{\pm1} \omega_{M}^{\mp\frac{1}{2}}\\
\partial_{\lambda}\mathcal{S}_{M}^{\pm\frac{3}{2}}&=&
\pm\frac{9}{8} k \omega_{M}^{\pm\frac{3}{2}}
\mp3 \mathcal{A}^{-1} \nu_{M}^{\pm\frac{1}{2}}
-\frac{9}{4} \mathcal{A}^0 \omega_{M}^{\pm\frac{3}{2}}
-\frac{3}{4} \mathcal{A}^{\pm1} \omega_{M}^{\pm\frac{1}{2}}
\pm\frac{3}{4} \upsilon^{\pm1} \mathcal{G}_{M}^{\pm\frac{1}{2}}
\pm3 \upsilon^{\pm2} \mathcal{G}_{M}^{\mp\frac{1}{2}}
+\frac{2}{3} \upsilon^0 \mathcal{S}_{M}^{\pm\frac{3}{2}}\nonumber\\
&+&2 \upsilon^{\pm1} \mathcal{S}_{M}^{\pm\frac{1}{2}}
+4 \upsilon^{\pm2} \mathcal{S}_{M}^{\mp\frac{1}{2}}
+\frac{3}{2} \phi^{\pm1} \mathcal{G}_{M}^{\pm\frac{1}{2}}
\pm\frac{9}{8} \mathcal{J} \omega_{M}^{\pm\frac{3}{2}}
-\varrho  \mathcal{S}_{M}^{\pm\frac{3}{2}}
\pm\frac{3}{2} \epsilon^0 \mathcal{S}_{M}^{\pm\frac{3}{2}}
\pm3 \epsilon^{\pm1} \mathcal{S}_{M}^{\pm\frac{1}{2}}
\pm\frac{1}{2} \phi^0 \mathcal{S}_{M}^{\pm\frac{3}{2}}\nonumber\\
&\pm&\phi^{\pm1} \mathcal{S}_{M}^{\pm\frac{1}{2}}
-\frac{3}{2} \mathcal{W}^{\pm1} \nu_{M}^{\pm\frac{1}{2}}
-\frac{3}{2} \mathcal{W}^{\pm2} \nu_{M}^{\mp\frac{1}{2}}
+\frac{9}{4} \mathcal{W}^0 \omega_{M}^{\pm\frac{3}{2}}
\mp\frac{3}{2} \mathcal{W}^{\pm1} \omega_{M}^{\pm\frac{1}{2}}
\mp\frac{3}{4} \mathcal{W}^{\pm2} \omega_{M}^{\mp\frac{1}{2}}\nonumber\\
&-&\frac{27}{4} \mathcal{L}^0 \omega_{M}^{\pm\frac{3}{2}}
-\frac{9}{4} \mathcal{L}^{\pm1} \omega_{M}^{\pm\frac{1}{2}}.\label{traqq2}
\end{eqnarray}
Analogously, for the chemical potentials the gauge transformations take form
\begin{eqnarray}
\partial_{\lambda}\eta&=&
k\partial_{\varphi}\varrho+\nu_{2}^{\frac{1}{2}} \nu_{1}^{-\frac{1}{2}}+\nu_{2}^{-\frac{1}{2}} \nu_{1}^{\frac{1}{2}}-\nu_{1}^{\frac{1}{2}} \nu_{2}^{-\frac{1}{2}}-\nu_{1}^{-\frac{1}{2}} \nu_{2}^{\frac{1}{2}}+\frac{4}{3} \psi_{2}^{\frac{3}{2}} \omega_{1}^{-\frac{3}{2}}+\frac{4}{3} \psi_{2}^{\frac{1}{2}} \omega_{1}^{-\frac{1}{2}}
+\frac{4}{3} \psi_{2}^{-\frac{1}{2}} \omega_{1}^{\frac{1}{2}}+\frac{4}{3} \psi_{2}^{-\frac{3}{2}} \omega_{1}^{\frac{3}{2}}\nonumber\\
&-&\frac{4}{3} \psi_{1}^{\frac{3}{2}} \omega_{2}^{-\frac{3}{2}}-\frac{4}{3} \psi_{1}^{\frac{1}{2}} \omega_{2}^{-\frac{1}{2}}-\frac{4}{3} \psi_{1}^{-\frac{1}{2}} \omega_{2}^{\frac{1}{2}}-\frac{4}{3} \psi_{1}^{-\frac{3}{2}} \omega_{2}^{\frac{3}{2}}, \\
\partial_{\lambda}\mu^0&=&
\frac{k}{4}  \partial_{\varphi}\epsilon^0-\frac{1}{2} \nu_{2}^{\frac{1}{2}} \nu_{1}^{-\frac{1}{2}}+\frac{1}{2} \nu_{2}^{-\frac{1}{2}} \nu_{1}^{\frac{1}{2}}-\frac{1}{2} \nu_{1}^{\frac{1}{2}} \nu_{2}^{-\frac{1}{2}}+\frac{1}{2} \nu_{1}^{-\frac{1}{2}} \nu_{2}^{\frac{1}{2}}-\frac{5}{8} \nu_{2}^{\frac{1}{2}} \omega_{1}^{-\frac{1}{2}}-\frac{5}{8} \nu_{2}^{-\frac{1}{2}} \omega_{1}^{\frac{1}{2}}\nonumber\\
&+&\frac{5}{8} \nu_{1}^{\frac{1}{2}} \omega_{2}^{-\frac{1}{2}}+\frac{5}{8} \nu_{1}^{-\frac{1}{2}} \omega_{2}^{\frac{1}{2}}-\frac{5}{3} \nu_{1}^{-\frac{1}{2}} \psi_{2}^{\frac{1}{2}}-\frac{5}{3} \nu_{1}^{\frac{1}{2}} \psi_{2}^{-\frac{1}{2}}+\frac{5}{3} \nu_{2}^{-\frac{1}{2}} \psi_{1}^{\frac{1}{2}}+\frac{5}{3} \nu_{2}^{\frac{1}{2}} \psi_{1}^{-\frac{1}{2}}+\frac{1}{2} \psi_{2}^{\frac{3}{2}} \omega_{1}^{-\frac{3}{2}}+\frac{1}{6} \psi_{2}^{\frac{1}{2}} \omega_{1}^{-\frac{1}{2}}\nonumber\\
&-&\frac{1}{6} \psi_{2}^{-\frac{1}{2}} \omega_{1}^{\frac{1}{2}}-\frac{1}{2} \psi_{2}^{-\frac{3}{2}} \omega_{1}^{\frac{3}{2}}+\frac{1}{2} \psi_{1}^{\frac{3}{2}} \omega_{2}^{-\frac{3}{2}}+\frac{1}{6} \psi_{1}^{\frac{1}{2}} \omega_{2}^{-\frac{1}{2}}-\frac{1}{6} \psi_{1}^{-\frac{1}{2}} \omega_{2}^{\frac{1}{2}}-\frac{1}{2} \psi_{1}^{-\frac{3}{2}} \omega_{2}^{\frac{3}{2}}-\xi^1 \phi^{-1}+\xi^{-1} \phi^1\nonumber\\
&+&\upsilon^2 \mathit{f}^{-2}+\frac{1}{2} \upsilon^1 \mathit{f}^{-1}-\frac{1}{2} \upsilon^{-1} \mathit{f}^1-\upsilon^{-2} \mathit{f}^2-\epsilon^{-1} \mu^1+\epsilon^1 \mu^{-1} \\
\partial_{\lambda}\mu^{\pm1}&=&
-\frac{k}{2} \partial_{\varphi} \epsilon^{\pm1}\pm\nu_{2}^{\pm\frac{1}{2}} \nu_{1}^{\pm\frac{1}{2}}\pm\nu_{1}^{\pm\frac{1}{2}} \nu_{2}^{\pm\frac{1}{2}}+\frac{15}{8} \nu_{2}^{\mp\frac{1}{2}} \omega_{1}^{\pm\frac{3}{2}}+\frac{5}{8} \nu_{2}^{\pm\frac{1}{2}} \omega_{1}^{\pm\frac{1}{2}}-\frac{15}{8} \nu_{\mp}^{\frac{1}{2}} \omega_{2}^{\pm\frac{3}{2}}-\frac{5}{8} \nu_{1}^{\pm\frac{1}{2}} \omega_{2}^{\pm\frac{1}{2}}\nonumber\\
&-&\frac{5}{3} \nu_{1}^{\pm\frac{1}{2}} \psi_{2}^{\pm\frac{1}{2}}
-\frac{5}{3} \nu_{1}^{\mp\frac{1}{2}} \psi_{2}^{\pm\frac{3}{2}}+\frac{5}{3} \nu_{2}^{\pm\frac{1}{2}} \psi_{1}^{\pm\frac{1}{2}}+\frac{5}{3} \nu_{2}^{\mp\frac{1}{2}} \psi_{1}^{\pm\frac{3}{2}}\mp\psi_{2}^{\mp\frac{1}{2}} \omega_{1}^{\pm\frac{3}{2}}\mp\frac{2}{3} \psi_{2}^{\pm\frac{1}{2}} \omega_{1}^{\pm\frac{1}{2}}\mp\frac{1}{3} \psi_{2}^{\pm\frac{3}{2}} \omega_{1}^{\mp\frac{1}{2}}\nonumber\\
&\mp&\psi_{1}^{\mp\frac{1}{2}} \omega_{2}^{\pm\frac{3}{2}}
+\frac{2}{3} \psi_{1}^{\pm\frac{1}{2}} \omega_{2}^{\pm\frac{1}{2}}
\mp\frac{1}{3} \psi_{1}^{\pm\frac{3}{2}} \omega_{2}^{\mp\frac{1}{2}}\pm2 \xi^0 \phi^{\pm1}\pm\xi^{\pm1} \phi^0-\frac{1}{2} \upsilon^{\mp1} \mathit{f}^{\pm2}\pm\upsilon^0 \mathit{f}^{\pm1}\pm\frac{3}{2} \upsilon^{\pm1} \mathit{f}^0\nonumber\\
&\pm&2 \upsilon^{\pm2} \mathit{f}^{\mp1}\pm2 \epsilon^{\pm1} \mu^0\pm\epsilon^0 \mu^{\pm1}\\
\partial_{\lambda}\xi^0&=&
\frac{k}{4}\partial_{\varphi}\phi_0+\frac{3}{8} \nu_{2}^{\frac{1}{2}} \omega _{1}^{-\frac{1}{2}}+\frac{3}{8} \nu_{2}^{-\frac{1}{2}} \omega _{1}^{\frac{1}{2}}-\frac{3}{8} \nu_{1}^{\frac{1}{2}} \omega _{2}^{-\frac{1}{2}}-\frac{3}{8} \nu_{1}^{-\frac{1}{2}} \omega_{2}^{\frac{1}{2}}+\nu_{1}^{-\frac{1}{2}} \psi_{2}^{\frac{1}{2}}+\nu _{1}^{\frac{1}{2}} \psi_{2}^{-\frac{1}{2}}-\nu _{2}^{-\frac{1}{2}} \psi_{1}^{\frac{1}{2}}\nonumber\\
&-&\nu _{2}^{\frac{1}{2}} \psi_{1}^{-\frac{1}{2}}
-\frac{3}{2} \psi_{2}^{\frac{3}{2}} \omega _{1}^{-\frac{3}{2}}-\frac{1}{2} \psi_{2}^{\frac{1}{2}} \omega _{1}^{-\frac{1}{2}}+\frac{1}{2} \psi_{2}^{-\frac{1}{2}} \omega _{1}^{\frac{1}{2}}+\frac{3}{2} \psi_{2}^{-\frac{3}{2}} \omega _{1}^{\frac{3}{2}}-\frac{3}{2} \psi_{1}^{\frac{3}{2}} \omega _{2}^{-\frac{3}{2}}-\frac{1}{2} \psi_{1}^{\frac{1}{2}} \omega _{2}^{-\frac{1}{2}}\nonumber\\
&+&\frac{1}{2} \psi_{1}^{-\frac{1}{2}} \omega _{2}^{\frac{1}{2}}
+\frac{3}{2} \psi_{1}^{-\frac{3}{2}} \omega _{2}^{\frac{3}{2}}
-\xi^1 \epsilon^{-1}+\xi^{-1} \epsilon^1+\upsilon^2 \mathit{f}^{-2}+\frac{1}{2} \upsilon^1 \mathit{f}^{-1}-\frac{1}{2} \upsilon^{-1} \mathit{f}^1-\upsilon^{-2} \mathit{f}^2\nonumber\\
&-&\phi^{-1} \mu^1+\phi^1 \mu^{-1} \\
\partial_{\lambda}\xi^{\pm1}&=&
-\frac{k}{2}  \partial_{\varphi}\phi^{\pm1}-\frac{9}{8} \nu_{2}^{\mp\frac{1}{2}} \omega_{1}^{\pm\frac{3}{2}}\pm\frac{3}{8} \nu_{2}^{\pm\frac{1}{2}} \omega_{1}^{\pm\frac{1}{2}}+\frac{9}{8} \nu_{1}^{\mp\frac{1}{2}} \omega_{2}^{\pm\frac{3}{2}}\mp\frac{3}{8} \nu_{1}^{\pm\frac{1}{2}} \omega_{2}^{\pm\frac{1}{2}}+\nu_{1}^{\pm\frac{1}{2}} \psi_{2}^{\pm\frac{1}{2}}+\nu_{1}^{\mp\frac{1}{2}} \psi_{2}^{\pm\frac{3}{2}}\nonumber\\
&-&\nu_{2}^{\pm\frac{1}{2}} \psi_{1}^{\pm\frac{1}{2}}
-\nu_{2}^{\mp\frac{1}{2}} \psi_{1}^{\pm\frac{3}{2}}\pm3 \psi_{2}^{\mp\frac{1}{2}} \omega_{1}^{\pm\frac{3}{2}}\pm2 \psi_{2}^{\pm\frac{1}{2}} \omega_{1}^{\pm\frac{1}{2}}\pm\psi_{2}^{\pm\frac{3}{2}} \omega_{1}^{\mp\frac{1}{2}}\pm3 \psi_{1}^{\mp\frac{1}{2}} \omega_{2}^{\pm\frac{3}{2}}\pm2 \psi_{1}^{\pm\frac{1}{2}} \omega_{2}^{\pm\frac{1}{2}}\nonumber\\
&\pm&\psi_{1}^{\pm\frac{3}{2}} \omega_{2}^{\mp\frac{1}{2}}\pm2 \xi^0 \epsilon^{\pm1}
\pm\xi^{\pm1} \epsilon^0\pm\frac{1}{2} \upsilon^{\mp1} \mathit{f}^{\pm2}\pm\upsilon^0 \mathit{f}^{\pm1}\pm\frac{3}{2} \upsilon^{\pm1} \mathit{f}^0\pm2 \upsilon^{\pm2} \mathit{f}^{\mp1}
\pm 2\phi^{\pm1} \mu^0\nonumber\\
&\pm&\phi^0 \mu^{\pm1}\\
\partial_{\lambda}\mathit{f}^0&=&
\frac{k}{3} \partial_{\varphi}\upsilon^0 +\frac{1}{2} \nu_{2}^{\frac{1}{2}} \omega_{1}^{-\frac{1}{2}}-\frac{1}{2} \nu_{2}^{-\frac{1}{2}} \omega_{1}^{\frac{1}{2}}+\frac{1}{2} \nu_{1}^{\frac{1}{2}} \omega_{2}^{-\frac{1}{2}}-\frac{1}{2} \nu_{1}^{-\frac{1}{2}} \omega_{2}^{\frac{1}{2}}+\frac{4}{3} \nu_{1}^{-\frac{1}{2}} \psi_{2}^{\frac{1}{2}}-\frac{4}{3} \nu_{1}^{\frac{1}{2}} \psi_{2}^{-\frac{1}{2}}+\frac{4}{3} \nu_{2}^{-\frac{1}{2}} \psi_{1}^{\frac{1}{2}}\nonumber\\
&-&\frac{4}{3} \nu_{2}^{\frac{1}{2}} \psi_{1}^{-\frac{1}{2}}+\frac{2}{3} \psi_{2}^{\frac{3}{2}} \omega_{1}^{-\frac{3}{2}}-\frac{2}{3} \psi_{2}^{\frac{1}{2}} \omega_{1}^{-\frac{1}{2}}-\frac{2}{3} \psi_{2}^{-\frac{1}{2}} \omega_{1}^{\frac{1}{2}}+\frac{2}{3} \psi_{2}^{-\frac{3}{2}} \omega_{1}^{\frac{3}{2}}-\frac{2}{3} \psi_{1}^{\frac{3}{2}} \omega_{2}^{-\frac{3}{2}}+\frac{2}{3} \psi_{1}^{\frac{1}{2}} \omega_{2}^{-\frac{1}{2}}+\frac{2}{3} \psi_{1}^{-\frac{1}{2}} \omega_{2}^{\frac{1}{2}}\nonumber\\
&-&\frac{2}{3} \psi_{1}^{-\frac{3}{2}} \omega_{2}^{\frac{3}{2}}-2 \xi^1 \upsilon^{-1}+2 \xi^{-1} \upsilon^1-2 \upsilon^{-1} \mu^1+2 \upsilon^1 \mu^{-1} -2 \epsilon^{-1} \mathit{f}^1+2 \epsilon^1 \mathit{f}^{-1}-2 \phi^{-1} \mathit{f}^1\nonumber\\
&+&2 \phi^1 \mathit{f}^{-1}
\end{eqnarray}
\begin{eqnarray}
\partial_{\lambda}\mathit{f}^{\pm1}&=&
-\frac{k}{2} \partial_{\varphi} \upsilon^{\pm1}\pm\frac{3}{4} \nu_{2}^{\mp\frac{1}{2}} \omega_{1}^{\pm\frac{3}{2}}\mp\frac{3}{4} \nu_{2}^{\pm\frac{1}{2}} \omega_{1}^{\pm\frac{1}{2}}\pm\frac{3}{4} \nu_{1}^{\mp\frac{1}{2}} \omega_{2}^{\pm\frac{3}{2}}\mp\frac{3}{4} \nu_{1}^{\pm\frac{1}{2}} \omega_{2}^{\pm\frac{1}{2}}\mp2 \nu_{1}^{\pm\frac{1}{2}} \psi_{2}^{\pm\frac{1}{2}}\pm\frac{2}{3} \nu_{1}^{\mp\frac{1}{2}} \psi_{2}^{\pm\frac{3}{2}}\nonumber\\
&\mp&2 \nu_{2}^{\pm\frac{1}{2}} \psi_{1}^{\pm\frac{1}{2}}\pm\frac{2}{3} \nu_{2}^{\mp\frac{1}{2}} \psi_{1}^{\pm\frac{3}{2}}\mp2 \psi_{2}^{\mp\frac{1}{2}} \omega_{1}^{\pm\frac{3}{2}}-\frac{2}{3} \psi_{2}^{\pm\frac{3}{2}} \omega_{1}^{\mp\frac{1}{2}}\pm2 \psi_{1}^{\mp\frac{1}{2}} \omega_{2}^{\pm\frac{3}{2}}+\frac{2}{3} \psi_{1}^{\pm\frac{3}{2}} \omega_{2}^{\mp\frac{1}{2}}\mp4 \xi^1 \upsilon^{\pm2}\nonumber\\
&\pm&2 \xi^0 \upsilon^{\pm1}\pm2 \xi^{\pm1} \upsilon^0\mp4 \upsilon^{\pm2} \mu^{\mp1}\pm2 \upsilon^0 \mu^{\pm1}\pm2 \upsilon^{\pm1} \mu^0 \pm3 \epsilon^{\pm1} \mathit{f}^0\pm\epsilon^0 \mathit{f}^{\pm1}+\epsilon^{\mp1} \mathit{f}^{\pm2}\nonumber\\
&\pm&3 \phi^{\pm1} \mathit{f}^0\pm\phi^0 \mathit{f}^{\pm1}
\mp\phi^{\mp1} \mathit{f}^{\pm2}
\end{eqnarray}
\begin{eqnarray}
\partial_{\lambda}\mathit{f}^{\pm2}&=&
2 k\partial_{\varphi} \upsilon^{\pm2}\pm3 \nu_{2}^{\pm\frac{1}{2}} \omega_{1}^{\pm\frac{3}{2}}\pm3 \nu_{1}^{\pm\frac{1}{2}} \omega_{2}^{\pm\frac{3}{2}}\mp\frac{8}{3} \nu_{1}^{\pm\frac{1}{2}} \psi_{2}^{\pm\frac{3}{2}}\mp\frac{8}{3} \nu_{2}^{\pm\frac{1}{2}} \psi_{1}^{\pm\frac{3}{2}}+4 \psi_{2}^{\pm\frac{1}{2}} \omega_{1}^{\pm\frac{3}{2}}\mp\frac{4}{3} \psi_{2}^{\pm\frac{3}{2}} \omega_{1}^{\pm\frac{1}{2}}\nonumber\\
&-&4 \psi_{1}^{\pm\frac{1}{2}} \omega_{2}^{\pm\frac{3}{2}}-\frac{4}{3} \psi_{1}^{\pm\frac{3}{2}} \omega_{2}^{\pm\frac{1}{2}}\mp16 \xi^0 \upsilon^{\pm2}\mp4 \xi^{\pm1} \upsilon^{\pm1}+16 \upsilon^{\pm2} \mu^0\mp4 \upsilon^{\pm1} \mu^{\pm1}\pm4 \epsilon^{\pm1} \mathit{f}^{\pm1}\nonumber\\
&\pm&2\epsilon^0 \mathit{f}^{\pm2}\pm4 \phi^{\pm1} \mathit{f}^{\pm1}\pm2 \phi^0 \mathit{f}^{\pm2} \\
\partial_{\lambda}\nu_{M}^{\pm\frac{1}{2}}&=&\mp k\partial_{\varphi}\nu_{M}^{\pm\frac{1}{2}}+\frac{10}{3} \xi^0 \nu_{M}^{\mp\frac{1}{2}}+\frac{10}{3} \xi^{\pm1} \nu_{M}^{\mp\frac{1}{2}}
\mp4 \xi^{\mp1} \omega_{M}^{\pm\frac{3}{2}}\mp\frac{8}{3} \xi^0 \omega_{M}^{\pm\frac{1}{2}}\mp\frac{4}{3} \xi^{\pm1} \omega_{M}^{\mp\frac{1}{2}}\mp\frac{8}{9} \upsilon^{\mp1} \psi_{M}^{\pm\frac{3}{2}}\nonumber\\
&\mp&\frac{16}{9} \upsilon^0 \psi_{M}^{\pm\frac{1}{2}}\mp\frac{8}{3} \upsilon^{\pm1} \psi_{M}^{\mp\frac{1}{2}}\mp\frac{32}{9} \upsilon^{\pm2} \psi_{M}^{\mp\frac{3}{2}}-\varrho  \nu_{M}^{\pm\frac{1}{2}}\pm\frac{1}{2} \epsilon^0 \nu_{M}^{\pm\frac{1}{2}}\pm\epsilon^{\pm1} \nu_{M}^{\mp\frac{1}{2}}\pm\frac{5}{6} \phi^0 \nu_{M}^{\pm\frac{1}{2}}\pm\frac{5}{3} \phi^{\pm1} \nu_{M}^{\mp\frac{1}{2}} \nonumber\\
&\mp&\eta \nu_{M}^{\pm\frac{1}{2}}\mp\frac{16}{9} \phi^{\mp1} \psi_{M}^{\pm\frac{3}{2}}-\frac{16}{9} \phi^0 \psi_{M}^{\pm\frac{1}{2}}-\frac{16}{9} \phi^{\pm1} \psi_{M}^{\mp\frac{1}{2}}-2
\mathit{f}^{\mp1} \omega_{M}^{\pm\frac{3}{2}}-2 \mathit{f}^0 \omega_{M}^{\pm\frac{1}{2}}-2 \mathit{f}^{\pm1} \omega_{M}^{\mp\frac{1}{2}} \nonumber\\
&-& 2\mathit{f}^{\pm2} \omega_{M}^{\mp\frac{3}{2}}+2 \mu^0 \nu_{M}^{\pm\frac{1}{2}}+2 \mu^{\pm1} \nu_{M}^{\mp\frac{1}{2}} \\
\partial_{\lambda}\psi_{M}^{\pm\frac{1}{2}}&=&\mp\frac{3}{8} k \omega_{M}^{\pm\frac{1}{2}}\mp2 \xi^0 \nu_{M}^{\pm\frac{1}{2}}\pm\xi^{\pm1} \nu_{M}^{\mp\frac{1}{2}}-\frac{3}{4} \xi^{\mp1} \omega_{M}^{\pm\frac{3}{2}}+\frac{1}{4} \xi^0 \omega_{M}^{\pm\frac{1}{2}}+\frac{1}{2} \xi^{\pm1} \omega_{M}^{\mp\frac{1}{2}}\mp\frac{1}{2} \upsilon^0 \nu_{M}^{\pm\frac{1}{2}}\nonumber\\
&\mp&\frac{3}{4} \upsilon^{\pm1} \nu_{M}^{\mp\frac{1}{2} }
-\frac{2}{3} \upsilon^{\mp1} \psi_{M}^{\pm\frac{3}{2}}-\frac{2}{3} \upsilon^0 \psi_{M}^{\pm\frac{1}{2}}+\frac{4}{3} \upsilon^{\pm2} \psi_{M}^{\mp\frac{3}{2}}-\frac{1}{2} \phi^0 \nu_{M}^{\pm\frac{1}{2}}+\frac{1}{2} \phi^{\pm1} \nu_{M}^{\mp\frac{1}{2}}\mp\frac{3}{8} \eta \omega_{M}^{\pm\frac{1}{2}} -\varrho  \psi_{M}^{\pm\frac{1}{2}}\nonumber\\
&\pm&\epsilon^{\mp1} \psi_{M}^{\pm\frac{3}{2}}\pm\frac{1}{2} \epsilon^0 \psi_{M}^{\pm\frac{1}{2}}
-2 \epsilon^{\pm1} \psi_{M}^{\mp\frac{1}{2}}\mp\frac{1}{3} \phi^{\mp1} \psi_{M}^{\pm\frac{3}{2}}\pm\frac{1}{6} \phi^0 \psi_{M}^{\pm\frac{1}{2}}\pm\frac{2}{3} \phi^{\pm1} \psi_{M}^{\mp\frac{1}{2}}\pm\frac{3}{2} \mathit{f}^0 \nu_{M}^{\pm\frac{1}{2}}-\frac{3}{2} \mathit{f}^{\pm1} \nu_{M}^{\mp\frac{1}{2}}\nonumber\\
&+&\frac{3}{2} \mathit{f}^{\mp1} \omega_{M}^{\pm\frac{3}{2}}\mp\frac{3}{4} \mathit{f}^0 \omega_{M}^{\pm\frac{1}{2}}
\pm\frac{3}{4} \mathit{f}^{\pm2} \omega_{M}^{\mp\frac{3}{2}}-\frac{9}{4} \mu^{\mp1} \omega_{M}^{\pm\frac{3}{2}}+\frac{3}{4} \mu^0 \omega_{M}^{\pm\frac{1}{2}}+\frac{3}{2} \mu^{\pm1} \omega_{M}^{\mp\frac{1}{2}}\\
\partial_{\lambda}\psi_{M}^{\pm\frac{3}{2}}&=&
\pm\frac{9}{8} k \omega_{M}^{\pm\frac{3}{2}}
\mp3 \xi^{-1} \nu_{M}^{\pm\frac{1}{2}}
-\frac{9}{4} \xi^0 \omega_{M}^{\pm\frac{3}{2}}
-\frac{3}{4} \xi^{\pm1} \omega_{M}^{\pm\frac{1}{2}}
\pm\frac{3}{4} \upsilon^{\pm1} \nu_{M}^{\pm\frac{1}{2}}
\pm3 \upsilon^{\pm2} \nu_{M}^{\mp\frac{1}{2}}
+\frac{2}{3} \upsilon^0 \psi_{M}^{\pm\frac{3}{2}}\nonumber\\
&+&2 \upsilon^{\pm1} \psi_{M}^{\pm\frac{1}{2}}
+4 \upsilon^{\pm2} \psi_{M}^{\mp\frac{1}{2}}
+\frac{3}{2} \phi^{\pm1} \nu_{M}^{\pm\frac{1}{2}}
\pm\frac{9}{8} \eta \omega_{M}^{\pm\frac{3}{2}}
-\varrho  \psi_{M}^{\pm\frac{3}{2}}
\pm\frac{3}{2} \epsilon^0 \psi_{M}^{\pm\frac{3}{2}}
\pm3 \epsilon^{\pm1} \psi_{M}^{\pm\frac{1}{2}}
\pm\frac{1}{2} \phi^0 \psi_{M}^{\pm\frac{3}{2}}\nonumber\\
&\pm&\phi^{\pm1} \psi_{M}^{\pm\frac{1}{2}}
-\frac{3}{2} \mathit{f}^{\pm1} \nu_{M}^{\pm\frac{1}{2}}
-\frac{3}{2} \mathit{f}^{\pm2} \nu_{M}^{\mp\frac{1}{2}}
+\frac{9}{4} \mathit{f}^0 \omega_{M}^{\pm\frac{3}{2}}
\mp\frac{3}{2} \mathit{f}^{\pm1} \omega_{M}^{\pm\frac{1}{2}}
\mp\frac{3}{4} \mathit{f}^{\pm2} \omega_{M}^{\mp\frac{1}{2}}\nonumber\\
&-&\frac{27}{4} \mu^0 \omega_{M}^{\pm\frac{3}{2}}
-\frac{9}{4} \mu^{\pm1} \omega_{M}^{\pm\frac{1}{2}}.
\end{eqnarray}
Following a similar approach as previous section, we act on to make out the canonical boundary charges $\mathcal{Q[\lambda]}$ that generates
the transformations $(\ref{traqq1})$-$(\ref{traqq2})$. As is well known, it is convenient to express the variation of the canonical boundary charge $\delta_\lambda \mathcal{Q}$ $(\ref{Qvar})$, to reach out the asymptotic symmetry algebra \cite{Banados:1998gg,Carlip:2005zn,Banados:1994tn,Banados:1998ta}. Hence, the canonical boundary charge $\mathcal{Q[\lambda]}$ can be obtained which reads
\begin{equation}\label{boundaryco10}
\mathcal{Q[\lambda]}=\int\mathrm{d}\varphi\;\left[
 \mathcal{J} \varrho
+\mathcal{L}^{i}\epsilon^{-i}
+\mathcal{A}^{i}\phi^{-i}
+\mathcal{W}^{i}\upsilon^{-i}
+\mathcal{G}_{M}^{p}\varsigma^{-p}
+\mathcal{S}_{M}^{p}\omega^{-p}
\right].
\end{equation}
The next step to derive the asymptotic symmetry algebra is to calculate Poisson bracket algebra using the standard method \cite{Blgojevic:2002}, which is acquired by the relation $(\ref{Qvar2})$ given for any phase space functional $\digamma$.

In light of the above statement, the operator product algebra  for the bosonic sector is  then achieved as
\begin{eqnarray}\label{affineope32}
\mathcal{L}^{i}(z_1)\mathcal{L}^{j}(z_2)\,& \sim &\,\frac{\frac{k}{2}\eta_{2}^{ij}}{z_{12}^{2}}\,+\,\frac{(i-j)}{z_{12}} \mathcal{L}^{i+j},\,\,\,\,\,
\mathcal{J}(z_1)\mathcal{J}(z_2)\, \sim  \frac{\frac{k}{2}\eta}{z_{12}^2},\\
\mathcal{L}^{i}(z_1)\mathcal{A}^{j}(z_2)\,& \sim &\, \frac{(i-j)}{z_{12}} \mathcal{A}^{i+j},\,\,\,\,\,
\mathcal{L}^{i}(z_1)\mathcal{W}^{j}(z_2)\,  \sim  \, \frac{(2i-j)}{z_{12}} \mathcal{W}^{i+j},\\
\mathcal{A}^{i}(z_1)\mathcal{A}^{j}(z_2)\, & \sim & \, \frac{\frac{k}{2}\eta_{2}^{ij}}{z_{12}^{2}}\,+\,\frac{(i-j)}{z_{12}} \mathcal{A}^{i+j} ,\,\,\,\,\,
\mathcal{A}^{i}(z_1)\mathcal{W}^{j}(z_2)\,  \sim \, \frac{(2i-j)}{z_{12}} \mathcal{A}^{i+j} ,\\
\mathcal{W}^{i}(z_1)\mathcal{W}^{j}(z_2)\,& \sim &\,\frac{\frac{k}{2}\eta_3^{ij}}{z_{12}^2}+\frac{1}{z_{12}}\left( {\frac{1}{3} (i-j) \left(2 i^2-i j+2 j^2-8\right) }\left(\mathcal{A}^{i+j}+\mathcal{L}^{i+j}\right)\right).
\end{eqnarray}
Furthermore, the explicit operator product algebra between the bosonic and fermionic sectors is given by
\begin{eqnarray}
\mathcal{J}(z_1)\mathcal{G}_{\pm}^{p}(z_2)\,& \sim &\,\pm\frac{\mathcal{G}_{\pm}^p}{z_{12}},\,\,\,\,\,
\mathcal{L}^{i}(z_1)\mathcal{G}_{\pm}^{p}(z_2)\,  \sim  \,\frac{({i\over 2}-p)}{z_{12}} \mathcal{G}_{\pm}^{i+p} ,\\
\mathcal{L}^{i}(z_1)\mathcal{S}_{\pm}^{p}(z_2)\,& \sim &\,\frac{({{3i}\over 2}-p)}{z_{12}} \mathcal{S}_{\pm}^{i+p} ,\,\,\,\,\,
\mathcal{J}(z_1)\mathcal{S}_{\pm}^{p}(z_2)\,  \sim  \,\pm\frac{\frac{k}{2}}{z_{12}}\mathcal{S}_{\pm}^{p+i} ,\\
\mathcal{G}_{\pm}^{p}(z_1)\mathcal{A}^{i}(z_2)\,& \sim &\,\mp\frac{\mathcal{S}_{\mp}^p}{z_{12}},\,\,\,\,\,
\mathcal{G}_{\pm}^{p}(z_1)\mathcal{W}^{i}(z_2)\, \sim \,-\frac{\frac{4}{3}({2i}-\frac{p}{2})}{z_{12}}\mathcal{S}_{\mp}^{p+i} ,\\
\mathcal{A}(z_1)\mathcal{S}_{\pm}^{p}(z_2)\,& \sim &\,\frac{1}{z_{12}} \left(\frac{1}{3}\left(\frac{3 i}{2}-p\right)\mathcal{S}_\pm^{i+p} \mp\frac{1}{4}\left(3 i^2-2 i p+p^2-\frac{9}{4}\right)\mathcal{G}_\pm^{i+p} \right),\\
\mathcal{S}_{\pm}^{p}(z_1)\mathcal{W}^{i}(z_2)\,& \sim &\,
\mp\frac{1}{z_{12}} \left(\frac{1}{3} \left(i^2-2 i p+2 p^2-\frac{5}{2}\right) \mathcal{S}_\mp^{p+i}
\right.\nonumber\\
&&\left.
+\frac{1}{8} \left(4 p^3-i^3+2 i^2 p-3 i p^2  -9 p+\frac{19}{4}i\right) \mathcal{G}_\mp^{p+i}\right).
\end{eqnarray}
Finally, the explicit operator product algebra for the fermionic sector yields
\begin{eqnarray}
\mathcal{G}_{\pm}^{p}(z_1)\mathcal{G}_{\pm}^{q}(z_2)\,& \sim &\, \frac{\frac{k}{2}\eta_{\frac{3}{2}}^{pq}}{z_{12}^{2}}\,+\,\frac{2}{z_{12}}\bigg( \mathcal{L}^{p+q} +\frac{5}{3}\mathcal{A}^{p+q} \pm\frac{(p-q)}{2} \mathcal{J}\bigg),\\
\mathcal{G}_{\pm}^{p}(z_1)\mathcal{S}_{\pm}^{q}(z_2)\,& \sim &\,\frac{2}{z_{12}} \left(\frac{3}{4} \mathcal{W}^{p+q}-\left(\frac{3 p}{2}-\frac{q}{2}\right) \mathcal{A}^{p+q}\right),\\
\mathcal{S}_{\pm}^{p}(z_1)\mathcal{S}_{\pm}^{q}(z_2)\,& \sim &\,  \frac{\frac{k}{2}\eta_{\frac{5}{2}}^{pq}}{z_{12}^{2}}\pm\frac{1}{z_{12}}\left(\frac{1}{8}\left(3 p^2-4 p q+3 q^2-\frac{9}{2}\right)\left(\mathcal{A}^{p+q}+3
\mathcal{L}^{p+q}\right)
\right.\nonumber\\
&&\left.
\mp\frac{3}{4}(p-q)\mathcal{W}^{p+q}
\pm\frac{3}{16}(p-q)\left(p^2+q^2-\frac{5}{2}\right)\mathcal{J}\right),
\end{eqnarray}
where $z_{12}=z_1-z_2$,\,or in the more compact form,
\begin{eqnarray}\label{ope111}
\mathfrak{\mathcal{\mathfrak{J}}}^{A}(z_1)\mathfrak{\mathcal{\mathfrak{J}}}^{B}(z_2)\,& \sim &\, \frac{\frac{k}{2}\eta^{AB}}{z_{12}^{2}}\,+\,\frac{\mathfrak{\mathcal{\mathfrak{f}}}^{AB}_{~~~C} \mathfrak{\mathcal{\mathfrak{J}}}^{C}(z_2)}{z_{12}}.
\end{eqnarray}
Note that $\eta^{AB}$ is the supertrace matrix and $\mathfrak{\mathcal{\mathfrak{f}}}^{AB}_{~~C}$'s are the structure constants of the related algebra with $(A,B=0,\pm1,\pm\frac{1}{2},0,\pm1,\pm\frac{1}{2},\pm\frac{3}{2})$,\,i.e,\,$\eta^{ip}=0$ and $\mathfrak{\mathcal{\mathfrak{f}}}^{ij}_{~~i+j}=(i-j)$.

Since the barred sector is completely analogous, the same results are obtained. Eventually, it follows that the asymptotic symmetry algebra for the loosest set of boundary conditions of $\mathcal{N}=(2,2)$ supergravity  is two copies of the affine $\mathfrak{sl}(3|2)_k$ algebra.
\subsection{For Superconformal Boundary}
As already discussed, it is convenient to point out that the super-conformal boundary conditions are the supersymmetric extension of the well-known Brown-Henneaux boundary conditions, presented in \cite{Brown:1986nw} for $AdS_3$ supergravity. In this section, our main goal is to construct the asymptotic symmetry algebra for the most general boundary conditions as the supersymmetric extension of the Brown-Henneaux boundary conditions. In accordance for this purpose, we launch into our section by imposing the Drinfeld-Sokolov heighest weight gauge condition on the $\mathfrak{sl}(3|2)$ Lie superalgebra valued connection (\ref{bouncondaf}), setting the fields as
\begin{eqnarray}
\mathcal{L}^0&=&\mathcal{A}^0=\mathcal{A}^{+1}=\mathcal{G}_{M}^{+\frac{1}{2}}=\mathcal{S}_{M}^{+\frac{1}{2}}=\mathcal{S}_{M}^{+\frac{3}{2}}=0,\nn\\
\,\mathcal{L}^{-1}&=&\mathcal{L},\,\mathcal{A}^{-1}=\mathcal{A},\,\mathcal{G}_{M}^{-\frac{1}{2}}=\mathcal{G}_{M},\,\mathcal{S}_{M}^{-\frac{3}{2}}=\mathcal{S}_{M},\,\gamma_{+1}\mathcal{L}^{+1}=1.
\end{eqnarray}
Correspondingly, the supersymmetric gauge connection is taken to be
\begin{eqnarray}\label{bouncondaysin}
a_\varphi &=&
 \Lt_{1}
+\gamma_{-1}\mathcal{L} \Lt_{-1}
+\vartheta_{-1}\mathcal{A} \At_{-1}
+\omega_{-2}\mathcal{W}\Wt_{-2}
+ \rho\mathcal{J}\Jt\nn\\
&+&\sigma_M^{-{\frac{1}{2}}}\mathcal{G}_{M}\Gt_{-{\frac{1}{2}}}^{M}
+\tau_M^{-{\frac{3}{2}}}\mathcal{S}_{M}\St_{-{\frac{3}{2}}}^{M}
,\\
a_t &=&
\eta\Jt
+\mu \Lt_{1}
+\xi \At_{1}
+\mathit{f} \Wt_{2}
+\nu_{M}\Gt_{{+\frac{1}{2}}}^{M}
+\psi_{M}\St_{{+\frac{3}{2}}}^{M}\nn\\
&+&\sum_{i=-1}^{0}\mu^i \Lt_{i}
+\sum_{i=-1}^{0}\xi^i \At_{i}
+\nu_{M}^{-{\frac{1}{2}}}\Gt_{-{\frac{1}{2}}}^{M}
+\sum_{p=-\frac{3}{2}}^{ \frac{1}{2}}\psi_{M}^{p}\St_{p}^{M},
\end{eqnarray}

where
$\eta$,\,
$\mu\equiv  \mu^{+1}$,\,
$\xi\equiv  \xi^{+1}$,\,
$\mathit{f}\equiv  \mathit{f}^{+2}$,\,
$\nu_{_M}\equiv \nu_{_M}^{+\frac{1}{2}} $,\, and
$\psi_{_M}\equiv \psi_{_M}^{+\frac{3}{2}} $
can be interpreted as the independent $chemical$ $potentials$.

When we request to gather up our next steps yielding the asymptotic symmetry algebra in a one paragraph, it is an appropriate option to give the following brief summary. The all functions except the $chemical$ $potentials$ can be fixed by the flattness conditions (\ref{flatness}) in the usual manner. The equations of motion for the fixed chemical potentials can also be obtained conventionally as the time evolution of the canonical boundary charges. But unfortunately, moving from the $\mathfrak{sl}(2|1)$-case to $\mathfrak{sl}(3|2)$-extension brings along the technical cumbersome although we have overcomen. Our preference is to ignore presenting these calculations here, because they take up too much space. We have maken choice to spare space for the calculations of the gauge parameter $\lambda$ and further.

In line with all these results to be obtained, we are now able to derive the superconformal asymptotic symmetry algebra. Using the Drinfeld-Sokolov reduction we have only six independent parameters as
$\mathcal{\varrho}$,\,
$\mathcal{\epsilon}\equiv\mathcal{\epsilon}^{+1}$,\,
$\mathcal{\phi}\equiv\mathcal{\phi}^{+1}$,\,
$\mathcal{\upsilon}\equiv\mathcal{\upsilon}^{+2}$,\,
$\mathcal{\varsigma}_{_M}\equiv\mathcal{\varsigma}_{_M}^{+{\frac{1}{2}}}$ and
$\mathcal{\omega}_{_M}\equiv\mathcal{\omega}_{_M}^{+{\frac{3}{2}}}$
It is now possible to compute the gauge transformations by considering all transformations $(\ref{boundarycond})$ that preserve the boundary conditions with the $\mathfrak{sl}(3|2)$ Lie superalgebra valued gauge parameter $\lambda$ which reads
\begin{eqnarray}
\lambda\,&=&\,b^{-1}\bigg[
 \varrho{\Jt}
+\epsilon {\Lt}_1
+\phi {\At}_1
+\upsilon {\Wt}_2
+\varsigma_-{\Gt}_{\frac{1}{2}}^{-}
+\varsigma_+{\Gt}_{\frac{1}{2}}^{+}
+\omega _-{\St}_{\frac{3}{2}}^{-}
+\omega_+{\St}_{\frac{3}{2}}^{+}
 \nonumber\\
&+&   \left(-\frac{\epsilon  \mathcal{G}_-}{2 k}-\frac{5 \phi  \mathcal{G}_-}{6 k}-\frac{20 \upsilon  \mathcal{S}_-}{3 k}-\frac{2 \mathcal{A} \omega _-}{k}+\frac{\mathcal{J} \varsigma_-}{2 k}-\varsigma_-'\right)
  {\Gt}_{-\frac{1}{2}}^{-}\nonumber\\
&+&\left(\frac{\epsilon  \mathcal{G}_+}{2 k}+\frac{5 \phi  \mathcal{G}_+}{6 k}-\frac{20 \upsilon  \mathcal{S}_+}{3 k}+\frac{2 \mathcal{A} \omega _+}{k}-\frac{\mathcal{J} \varsigma_+}{2
   k}-\varsigma_+'\right){\Gt}_{-\frac{1}{2}}^{+}\nonumber\\
&+&\frac{1}{4}{\Lt}_0 \left(-\frac{15 \omega _- \mathcal{G}_+}{2 k}-\frac{15 \mathcal{G}_- \omega _+}{2 k}-4 \epsilon '\right)-{\Wt}_1 \upsilon '+\frac{1}{4}
  {\At}_0 \left(\frac{9 \omega _- \mathcal{G}_+}{2 k}+\frac{9 \mathcal{G}_- \omega _+}{2 k}-4 \phi '\right)\nonumber\\
&+&{\St}_{\frac{1}{2}}^{-} \left(\frac{4 \upsilon  \mathcal{G}_-}{3 k}+\frac{\mathcal{J} \omega _-}{2 k}-\omega
   _-'\right)+{\St}_{\frac{1}{2}}^{+} \left(-\frac{4 \upsilon  \mathcal{G}_+}{3 k}-\frac{\mathcal{J} \omega _+}{2 k}-\omega _+'\right)\nonumber\\
&+&\frac{1}{8}{\Lt}_{-1} \left(\frac{8 \mathcal{L} \epsilon }{k}-\frac{80 \mathcal{W}
   \upsilon }{k}+\frac{8 \mathcal{A} \phi }{k}-\frac{5 \mathcal{J} \omega _- \mathcal{G}_+}{2 k^2}-\frac{4 \varsigma_- \mathcal{G}_+}{k}+\frac{15 \omega _- \mathcal{S}_+}{k}+\frac{5 \mathcal{J} \mathcal{G}_- \omega _+}{2
   k^2}
 \right.\nonumber\\
&+&\left.
\frac{15 \mathcal{S}_- \omega _+}{k}
+\frac{4 \mathcal{G}_- \varsigma_+}{k}+\frac{15 \omega _+ \mathcal{G}_-'}{2 k}+\frac{25 \mathcal{G}_+ \omega _-'}{2 k}+\frac{15 \omega _- \mathcal{G}_+'}{2 k}+\frac{25
   \mathcal{G}_- \omega _+'}{2 k}+4 \epsilon ''\right)\nonumber\\
&+&\frac{1}{4}{\Wt}_0 \left(\frac{8 \mathcal{A} \upsilon }{k}+\frac{8 \mathcal{L} \upsilon }{k}+\frac{3 \omega _- \mathcal{G}_+}{2 k}-\frac{3 \mathcal{G}_- \omega
   _+}{2 k}+2 \upsilon ''\right)+\frac{1}{8}{\At}_{-1} \left(\frac{8 \mathcal{A} \epsilon }{k} -\frac{80 \mathcal{W} \upsilon }{k}
 \right.\nonumber\\
&+&\left.
  \frac{8 \mathcal{L} \phi }{k}+\frac{3 \mathcal{J} \omega _- \mathcal{G}_+}{2
   k^2}-\frac{45 \omega _- \mathcal{S}_+}{k}-\frac{3 \mathcal{J} \mathcal{G}_- \omega _+}{2 k^2}-\frac{45 \mathcal{S}_- \omega _+}{k}-\frac{9 \omega _+ \mathcal{G}_-'}{2 k}-\frac{15 \mathcal{G}_+ \omega _-'}{2 k}
  \right.\nonumber\\
&-&\left.
\frac{9 \omega _- \mathcal{G}_+'}{2 k}-\frac{15 \mathcal{G}_- \omega _+'}{2 k}+4 \phi ''\right)+\frac{1}{6}{\St}_{-\frac{1}{2}}^{-} \left(\frac{3 \omega _- \mathcal{J}^2}{4 k^2}+\frac{2 \upsilon  \mathcal{G}_-
   \mathcal{J}}{k^2}-\frac{3 \omega _-' \mathcal{J}}{k}+\frac{2 \phi  \mathcal{G}_-}{k}
  \right.\nonumber\\
&-&\left.
 \frac{20 \upsilon  \mathcal{S}_-}{k}+\frac{3 \mathcal{A} \omega _-}{k}+\frac{9 \mathcal{L} \omega _-}{k}-\frac{3 \omega _-
   \mathcal{J}'}{2 k}-\frac{7 \mathcal{G}_- \upsilon '}{k}-\frac{4 \upsilon  \mathcal{G}_-'}{k}+3 \omega _-''\right)\nonumber\\
&+&\frac{1}{6}{\St}_{-\frac{1}{2}}^{+} \left(\frac{3 \omega _+ \mathcal{J}^2}{4 k^2}+\frac{2 \upsilon
   \mathcal{G}_+ \mathcal{J}}{k^2}+\frac{3 \omega _+' \mathcal{J}}{k}+\frac{2 \phi  \mathcal{G}_+}{k}+\frac{20 \upsilon  \mathcal{S}_+}{k}+\frac{3 \mathcal{A} \omega _+}{k}+\frac{9 \mathcal{L} \omega _+}{k}
 \right.\nonumber\\
&+&\left.
\frac{3 \omega_+ \mathcal{J}'}{2 k}+\frac{7 \mathcal{G}_+ \upsilon '}{k}+\frac{4 \upsilon  \mathcal{G}_+'}{k}+3 \omega _+''\right)+\frac{1}{12}{\Wt}_{-1} \left(\frac{8 \upsilon  \mathcal{G}_- \mathcal{G}_+}{k^2}+\frac{3
   \mathcal{J} \omega _- \mathcal{G}_+}{2 k^2}-\frac{9 \omega _-' \mathcal{G}_+}{2 k}
 \right.\nonumber\\
&+&\left.
   \frac{15 \omega _- \mathcal{S}_+}{k}+\frac{3 \mathcal{J} \mathcal{G}_- \omega _+}{2 k^2}-\frac{15 \mathcal{S}_- \omega _+}{k}-\frac{8
   \upsilon  \mathcal{A}'}{k}-\frac{8 \upsilon  \mathcal{L}'}{k}-\frac{20 \mathcal{A} \upsilon '}{k}-\frac{20 \mathcal{L} \upsilon '}{k}+\frac{3 \omega _+ \mathcal{G}_-'}{2 k}
  \right.\nonumber\\
 &-&\left.
\frac{3 \omega _- \mathcal{G}_+'}{2 k}+\frac{9
   \mathcal{G}_- \omega _+'}{2 k}-2 \upsilon ^{(3)}\right)+\frac{1}{54}{\St}_{-\frac{3}{2}}^{-} \left(\frac{9 \omega _- \mathcal{J}^3}{8 k^3}+\frac{3 \upsilon  \mathcal{G}_- \mathcal{J}^2}{k^3}-\frac{27 \omega _-'
   \mathcal{J}^2}{4 k^2}+\frac{3 \phi  \mathcal{G}_- \mathcal{J}}{k^2}
  \right.\nonumber\\
&-&\left.
\frac{30 \upsilon  \mathcal{S}_- \mathcal{J}}{k^2}+\frac{21 \mathcal{A} \omega _- \mathcal{J}}{2 k^2}+\frac{63 \mathcal{L} \omega _- \mathcal{J}}{2
   k^2}-\frac{27 \omega _- \mathcal{J}' \mathcal{J}}{4 k^2}-\frac{33 \mathcal{G}_- \upsilon ' \mathcal{J}}{2 k^2}-\frac{12 \upsilon  \mathcal{G}_-' \mathcal{J}}{k^2}+\frac{27 \omega _-'' \mathcal{J}}{2 k}
 \right.\nonumber
  \end{eqnarray}
\begin{eqnarray}
&+&\left.
   \frac{40   \mathcal{A} \upsilon  \mathcal{G}_-}{k^2}+\frac{72 \mathcal{L} \upsilon  \mathcal{G}_-}{k^2}-\frac{90 \epsilon  \mathcal{S}_-}{k}-\frac{30 \phi  \mathcal{S}_-}{k}-\frac{180 \mathcal{W} \omega _-}{k}+\frac{24 \mathcal{A}   \varsigma_-}{k}+\frac{18 \mathcal{G}_- \omega _- \mathcal{G}_+}{k^2}
 \right.\nonumber\\
&+&\left.
   \frac{9 \mathcal{G}_-^2 \omega _+}{k^2}-\frac{9 \omega _- \mathcal{A}'}{k}-\frac{6 \upsilon  \mathcal{G}_- \mathcal{J}'}{k^2}-\frac{27 \omega _-   \mathcal{L}'}{k}+\frac{120 \mathcal{S}_- \upsilon '}{k}-\frac{18 \mathcal{G}_- \phi '}{k}-\frac{6 \phi  \mathcal{G}_-'}{k}+\frac{33 \upsilon ' \mathcal{G}_-'}{k}
 \right.\nonumber\\
&+&\left.
\frac{60 \upsilon  \mathcal{S}_-'}{k}-\frac{21 \mathcal{A}   \omega _-'}{k}-\frac{63 \mathcal{L} \omega _-'}{k}+\frac{27 \mathcal{J}' \omega _-'}{2 k}+\frac{9 \omega _- \mathcal{J}''}{2 k}+\frac{27 \mathcal{G}_- \upsilon ''}{k}+\frac{12 \upsilon  \mathcal{G}_-''}{k}-9 \omega_-{}^{(3)}\right)\nonumber\\
&+&\frac{1}{54}{\St}_{-\frac{3}{2}}^{+} \left(-\frac{9 \omega _+ \mathcal{J}^3}{8 k^3}-\frac{3 \upsilon  \mathcal{G}_+ \mathcal{J}^2}{k^3}-\frac{27 \omega _+' \mathcal{J}^2}{4 k^2}-\frac{3 \phi
   \mathcal{G}_+ \mathcal{J}}{k^2}-\frac{30 \upsilon  \mathcal{S}_+ \mathcal{J}}{k^2}-\frac{21 \mathcal{A} \omega _+ \mathcal{J}}{2 k^2}
  \right.\nonumber\\
&-&\left.
\frac{63 \mathcal{L} \omega _+ \mathcal{J}}{2 k^2}-\frac{27 \omega _+ \mathcal{J}'
   \mathcal{J}}{4 k^2}-\frac{33 \mathcal{G}_+ \upsilon ' \mathcal{J}}{2 k^2}-\frac{12 \upsilon  \mathcal{G}_+' \mathcal{J}}{k^2}-\frac{27 \omega _+'' \mathcal{J}}{2 k}+\frac{9 \omega _- \mathcal{G}_+^2}{k^2}-\frac{40
   \mathcal{A} \upsilon  \mathcal{G}_+}{k^2}
 \right.\nonumber\\
&-&\left.
\frac{72 \mathcal{L} \upsilon  \mathcal{G}_+}{k^2}-\frac{90 \epsilon  \mathcal{S}_+}{k}-\frac{30 \phi  \mathcal{S}_+}{k}+\frac{180 \mathcal{W} \omega _+}{k}+\frac{18
   \mathcal{G}_- \mathcal{G}_+ \omega _+}{k^2}-\frac{24 \mathcal{A} \varsigma_+}{k}-\frac{9 \omega _+ \mathcal{A}'}{k}
 \right.\nonumber\\
&-&\left.
\frac{6 \upsilon  \mathcal{G}_+ \mathcal{J}'}{k^2}-\frac{27 \omega _+ \mathcal{L}'}{k}-\frac{120
   \mathcal{S}_+ \upsilon '}{k}-\frac{18 \mathcal{G}_+ \phi '}{k}-\frac{6 \phi  \mathcal{G}_+'}{k}-\frac{33 \upsilon ' \mathcal{G}_+'}{k}-\frac{60 \upsilon  \mathcal{S}_+'}{k}-\frac{21 \mathcal{A} \omega _+'}{k}
 \right.\nonumber\\
&-&\left.
\frac{63\mathcal{L} \omega _+'}{k}-\frac{27 \mathcal{J}' \omega _+'}{2 k}-\frac{9 \omega _+ \mathcal{J}''}{2 k}-\frac{27 \mathcal{G}_+ \upsilon ''}{k}-\frac{12 \upsilon  \mathcal{G}_+''}{k}-9 \omega
   _+{}^{(3)}\right)+\frac{1}{48}{\Wt}_{-2} \left(\frac{48 \upsilon  \mathcal{A}^2}{k^2}
 \right.\nonumber\\
&+&\left.
   \frac{96 \mathcal{L} \upsilon  \mathcal{A}}{k^2}+\frac{27 \omega _- \mathcal{G}_+ \mathcal{A}}{2 k^2}-\frac{27 \mathcal{G}_-
   \omega _+ \mathcal{A}}{2 k^2}+\frac{32 \upsilon '' \mathcal{A}}{k}+\frac{120 \mathcal{W} \epsilon }{k}+\frac{48 \mathcal{L}^2 \upsilon }{k^2}+\frac{120 \mathcal{W} \phi }{k}
 \right.\nonumber\\
&+&\left.
\frac{10 \upsilon  \mathcal{S}_-\mathcal{G}_+}{k^2}+\frac{9 \mathcal{J}^2 \omega _- \mathcal{G}_+}{8 k^3}+\frac{45 \mathcal{L} \omega _- \mathcal{G}_+}{2 k^2}+\frac{10 \upsilon  \mathcal{G}_- \mathcal{S}_+}{k^2}+\frac{15 \mathcal{J} \omega _-\mathcal{S}_+}{k^2}-\frac{30 \varsigma_- \mathcal{S}_+}{k}
 \right.\nonumber\\
&-&\left.
\frac{9 \mathcal{J}^2 \mathcal{G}_- \omega _+}{8 k^3}-\frac{45 \mathcal{L} \mathcal{G}_- \omega _+}{2 k^2}+\frac{15 \mathcal{J} \mathcal{S}_- \omega_+}{k^2}-\frac{30 \mathcal{S}_- \varsigma_+}{k}-\frac{15 \omega _- \mathcal{G}_+ \mathcal{J}'}{4 k^2}-\frac{15 \mathcal{G}_- \omega _+ \mathcal{J}'}{4 k^2}
  \right.\nonumber\\
&-&\left.
\frac{29 \mathcal{G}_- \mathcal{G}_+ \upsilon '}{k^2}+\frac{28\mathcal{A}' \upsilon '}{k}+\frac{28 \mathcal{L}' \upsilon '}{k}-\frac{14 \upsilon  \mathcal{G}_+ \mathcal{G}_-'}{k^2}-\frac{3 \mathcal{J} \omega _+ \mathcal{G}_-'}{2 k^2}+\frac{15 \omega _+ \mathcal{S}_-'}{k}-\frac{6\mathcal{J} \mathcal{G}_+ \omega _-'}{k^2}
 \right.\nonumber\\
&-&\left.
\frac{45 \mathcal{S}_+ \omega _-'}{k}-\frac{14 \upsilon  \mathcal{G}_- \mathcal{G}_+'}{k^2}-\frac{3 \mathcal{J} \omega _- \mathcal{G}_+'}{2 k^2}+\frac{6 \omega _-'
   \mathcal{G}_+'}{k}-\frac{15 \omega _- \mathcal{S}_+'}{k}-\frac{6 \mathcal{J} \mathcal{G}_- \omega _+'}{k^2}+\frac{45 \mathcal{S}_- \omega _+'}{k}
  \right.\nonumber\\
&-&\left.
\frac{6 \mathcal{G}_-' \omega _+'}{k}+\frac{8 \upsilon
   \mathcal{A}''}{k}+\frac{8 \upsilon  \mathcal{L}''}{k}+\frac{32 \mathcal{L} \upsilon ''}{k}-\frac{3 \omega _+ \mathcal{G}_-''}{2 k}+\frac{9 \mathcal{G}_+ \omega _-''}{k}+\frac{3 \omega _- \mathcal{G}_+''}{2 k}-\frac{9
   \mathcal{G}_- \omega _+''}{k}+2 \upsilon ^{(4)}\right)
\bigg]b.
   \nonumber\\
\end{eqnarray}
Since we are dealing with the asymptotic symmetries, it is natural to demand obtaining the canonical boundary charges $\mathcal{Q[\lambda]}$.
So, the variation of the canonical boundary charge, i.e., $\delta_\lambda \mathcal{Q}$ $(\ref{Qvar})$ can be integrated to yield
\begin{equation}
\mathcal{Q[\lambda]}=\int\mathrm{d}\varphi\;
\left[
 \mathcal{J}\varrho
+\mathcal{L}\epsilon
+\mathcal{A}\phi
+\mathcal{W}\upsilon
+\mathcal{G}_{_M}\varsigma_{_M}
+\mathcal{S}_{_M}\omega_{_M}
\right].
\label{boundaryco11111}
\end{equation}
But,\,these canonical boundary charges do not give a convenient asymptotic operator product algebra in the complex coordinates by using $(\ref{Qvar2})$ for $\mathcal{N}=(2,2)$ superconformal boundary,
\begin{eqnarray}
\mathcal{L}(z_1)\mathcal{L}(z_2)&\sim&\frac{3 k}{z_{12}^4}+\frac{2 \mathcal{L}}{z_{12}^2}+\frac{\mathcal{L}'-\frac{\mathcal{G}_{+} \mathcal{G}_{-}}{k}}{z_{12}}\\
\mathcal{L}(z_1)\mathcal{J}(z_2)&\sim&0\\
\mathcal{L}(z_1)\mathcal{G}_{\pm}(z_2)&\sim&\frac{3 \mathcal{G}_{+}}{2 z_{12}^2}+\frac{\mathcal{G}_{+}'\pm\frac{ \mathcal{J}\mathcal{G}_{+}}{2 k}}{z_{12}}
\end{eqnarray}
\begin{eqnarray}
\mathcal{L}(z_1)\mathcal{A}(z_2)&\sim&\frac{\mathcal{A}'}{z_{12}}+\frac{2 \mathcal{A}}{z_{12}^2}\\
\mathcal{L}(z_1)\mathcal{S}_{\pm}(z_2)&\sim&\frac{5 \mathcal{S}_{+}}{2 z_{12}^2}+\frac{\mathcal{S}_{+}'\pm\frac{\mathcal{J} \mathcal{S}_{+}}{2 k}}{z_{12}}\\
\mathcal{L}(z_1)\mathcal{W}(z_2)&\sim&+\frac{3 \mathcal{W}}{z_{12}^2}+\frac{1}{z_{12}}\bigg(\mathcal{W}'+\frac{\mathcal{G}_{+} \mathcal{S}_{-}}{k}-\frac{\mathcal{S}_{+} \mathcal{G}_{-}}{k}\bigg)\\
\mathcal{J}(z_1)\mathcal{J}(z_2)&\sim&\frac{2 k}{z_{12}^2}\\
\mathcal{J}(z_1)\mathcal{G}_{\pm}(z_2)&\sim&\mp\frac{\mathcal{G}_{\pm}}{z_{12}}\\
\mathcal{J}(z_1)\mathcal{A}(z_2)&\sim&0\\
\mathcal{J}(z_1)\mathcal{S}_{\pm}(z_2)&\sim&\mp\frac{\mathcal{S}_{\pm}}{z_{12}}\\
\mathcal{J}(z_1)\mathcal{W}(z_2)&\sim&0\\
\mathcal{G}_{\pm}(z_1)\mathcal{G}_{\pm}(z_2)\,&\sim & \,{\mp{4k}\over{z_{12}^{3}}}\,-\,{{2\mathcal{J}}\over{z_{12}^{2}}}\,\, + \,\frac{\mp2}{z_{12}}\bigg( \mathcal{L}+\frac{\mathcal{J}\mathcal{J}}{4k}\pm\frac{\mathcal{J}'}{2}\mp\frac{10 \mathcal{A}}{3}\bigg)\\
\mathcal{G}_{\pm}(z_1)\mathcal{G}_{\mp}(z_2)&\sim&0\\
\mathcal{G}_{\pm}(z_1)\mathcal{A}(z_2)&\sim&\mp\frac{15 \mathcal{S}_{\pm}}{4 z_{12}}\\
\mathcal{G}_{\pm}(z_1)\mathcal{S}_{\pm}(z_2)&\sim&\pm\frac{16 \mathcal{A}}{15 z_{12}^2}\pm\frac{1}{z_{12}}\bigg(\frac{4 \mathcal{A}'}{15}\pm\frac{8 \mathcal{A} \mathcal{J}}{15 k}\mp4 \mathcal{W}\bigg)\\
\mathcal{G}_{\pm}(z_1)\mathcal{S}_{\mp}(z_2)&\sim&0\\
\mathcal{G}_{\pm}(z_1)\mathcal{W}(z_2)&\sim&-\frac{5 \mathcal{S}_{\mp}}{4 z_{12}^2}-\frac{1}{z_{12}}\bigg(\frac{\mathcal{S}_{\mp}'}{4}-\frac{2 \mathcal{A} \mathcal{G}_{\mp}}{15 k}\mp\frac{\mathcal{J} \mathcal{S}_{\mp}}{2 k}
\bigg)\\
\mathcal{A}(z_1)\mathcal{A}(z_2)&\sim&\frac{3 k}{z_{12}^4}+\frac{2 \mathcal{L}}{z_{12}^2}+\frac{\mathcal{L}'-\frac{\mathcal{G}_{+} \mathcal{G}_{-}}{4 k}}{z_{12}}\\
\mathcal{S}_{\pm}(z_1)\mathcal{S}_{\mp}(z_2)&\sim&-\frac{2 \mathcal{G}_{\mp} \mathcal{G}_{\mp}'}{5 k z_{12}}\\
\mathcal{A}(z_1)\mathcal{S}_{\pm}(z_2)&\sim&
\frac{4 \mathcal{G}_+}{5 z_{12}^3}+\frac{1}{z_{12}^2}\left(\frac{5 \mathcal{S}_\pm}{6}+\frac{4 \mathcal{G}_\pm'}{15}\pm\frac{2 \mathcal{G}_\pm \mathcal{J}}{15 k}\right)\nonumber\\
&+&\frac{1}{z_{12}}\left(\frac{\mathcal{G}_\pm \mathcal{J}^2}{60 k^2}-\frac{3 \mathcal{A} \mathcal{G}_\pm}{5 k}\pm\frac{\mathcal{G}_\pm \mathcal{J}'}{30 k}\pm\frac{\mathcal{J} \mathcal{G}_\pm'}{15   k}+\frac{3 \mathcal{G}_\pm \mathcal{L}}{5 k}\pm\frac{\mathcal{J} \mathcal{S}_\pm}{6 k}+\frac{\mathcal{S}_\pm'}{3}+\frac{\mathcal{G}_\pm''}{15}\right)\\
\mathcal{W}(z_1)\mathcal{W}(z_2)&\sim&\frac{2 k}{z_{12}^6}+\frac{2(\mathcal{A}+\mathcal{L})}{z_{12}^4}-\frac{1}{z_{12}^3}\left(\mathcal{A}'+\frac{13 \mathcal{G}_- \mathcal{G}_+}{12 k}+\mathcal{L}'\right)\nonumber\\
&+&\frac{1}{z_{12}^2}\left(\frac{3 \mathcal{A}''}{10}-\frac{13 \mathcal{G}_+ \mathcal{G}_-'}{24 k}
-\frac{13 \mathcal{G}_- \mathcal{G}_+'}{24 k}+\frac{16 \mathcal{A}^2}{15 k z_{12}^2}+\frac{32 \mathcal{A} \mathcal{L}}{15 k z_{12}^2}+\frac{16 \mathcal{L}^2}{15 k}+\frac{3 \mathcal{L}''}{10}\right)\nonumber\\
&-&\frac{1}{z_{12}}\left(\frac{\mathcal{G}_- \mathcal{G}_+ \mathcal{J}^2}{30 k^3}+\frac{32 \mathcal{A} \mathcal{G}_- \mathcal{G}_+}{45 k^2}-\frac{\mathcal{A}^{(3)}}{15}-\frac{\mathcal{G}_+ \mathcal{J}
   \mathcal{S}_-}{3 k^2}+\frac{\mathcal{G}_- \mathcal{J} \mathcal{S}_+}{3 k^2}-\frac{\mathcal{G}_+ \mathcal{J} \mathcal{G}_-'}{15 k^2}
\right.\nonumber\\
&+&\left.
\frac{\mathcal{G}_- \mathcal{J} \mathcal{G}_+'}{15 k^2}+\frac{16 \mathcal{G}_-   \mathcal{G}_+ \mathcal{L}}{15 k^2}-\frac{16 \mathcal{L} \mathcal{A}'}{15 k}-\frac{16 \mathcal{A} \mathcal{A}'}{15 k}-\frac{16 \mathcal{A} \mathcal{L}'}{15 k}-\frac{5 \mathcal{S}_+ \mathcal{G}_-'}{12 k}+\frac{\mathcal{G}_+   \mathcal{S}_-'}{4 k}
\right.\nonumber\\
&-&\left.
\frac{5 \mathcal{S}_- \mathcal{G}_+'}{12 k}+\frac{\mathcal{G}_- \mathcal{S}_+'}{4 k}+\frac{7 \mathcal{G}_-' \mathcal{G}_+'}{30 k}+\frac{11 \mathcal{G}_+ \mathcal{G}_-''}{60 k}+\frac{11 \mathcal{G}_-   \mathcal{G}_+''}{60 k}+\frac{10 \mathcal{S}_- \mathcal{S}_+}{k}-\frac{16 \mathcal{L} \mathcal{L}'}{15 k}-\frac{\mathcal{L}^{(3)}}{15}\right)\nonumber\\
\end{eqnarray}
\begin{eqnarray}
\mathcal{S}_{\pm}(z_1)\mathcal{W}(z_2)&\sim&\mp\frac{3 \mathcal{G}_\mp}{4 z_{12}^4}\pm\frac{1}{z_{12}^3}\left(\frac{5 \mathcal{S}_\mp}{4}\mp\frac{\mathcal{G}_\mp \mathcal{J}}{4 k}-\frac{\mathcal{G}_\mp'}{4}\right)\nonumber\\
&\mp&\frac{1}{z_{12}^2}\left(\frac{\mathcal{G}_\mp''}{16}+\frac{3 \mathcal{G}_\mp \mathcal{J}^2}{64 k^2}+\frac{91 \mathcal{A} \mathcal{G}_\mp}{240 k}\pm\frac{5 \mathcal{G}_\mp\mathcal{J}'}{32 k}\pm\frac{\mathcal{J}
   \mathcal{G}_\mp'}{16 k}+\frac{11 \mathcal{G}_\mp \mathcal{L}}{16 k}\mp\frac{3 \mathcal{J} \mathcal{S}_\mp}{8 k}-\frac{\mathcal{S}_-'}{2}\right)\nonumber\\
&-&\frac{1}{z_{12}}\left(\frac{\mathcal{G}_\mp\mathcal{J}^3}{160 k^3}\mp\frac{\mathcal{G}_\mp{}^{(3)}}{80}+\frac{13 \mathcal{A} \mathcal{G}_\mp \mathcal{J}}{120 k^2}\pm\frac{3 \mathcal{J}^2 \mathcal{G}_-'}{320 k^2}\pm\frac{9 \mathcal{G}_\mp \mathcal{J} \mathcal{J}'}{160 k^2}\pm\frac{9 \mathcal{G}_\mp \mathcal{J}\mathcal{L}}{40 k^2}\mp\frac{\mathcal{J}^2 \mathcal{S}_\mp}{16 k^2}
\right.\\
&\pm&\left.
\frac{13 \mathcal{G}_\mp \mathcal{A}'}{80 k}\pm\frac{13 \mathcal{A} \mathcal{G}_\mp'}{80 k}\mp\frac{11 \mathcal{A} \mathcal{S}_\mp}{4 k}+\frac{9 \mathcal{G}_\mp\mathcal{J}''}{160 k}+\frac{7 \mathcal{G}_\mp' \mathcal{J}'}{160 k}+\frac{\mathcal{J} \mathcal{G}_\mp''}{80 k}-\frac{5 \mathcal{G}_\mp \mathcal{W}}{2 k}\pm\frac{27 \mathcal{G}_\mp \mathcal{L}'}{80 k}
\right.\nonumber\\
&\pm&\left.
\frac{19 \mathcal{L} \mathcal{G}_\mp'}{80 k}-\frac{\mathcal{S}_\mp \mathcal{J}'}{4 k}-\frac{\mathcal{J} \mathcal{S}_\mp'}{8 k}\mp\frac{5 \mathcal{S}_\mp \mathcal{L}}{4
   k}\mp\frac{\mathcal{S}_\mp''}{8}\right)
\end{eqnarray}
\begin{eqnarray}
\mathcal{S}_{\pm}(z_1)\mathcal{S}_{\pm}(z_2)&\sim&\frac{12 k}{5 z_{12}^5}\pm\frac{6 \mathcal{J}}{5 z_{12}^4}+\frac{1}{z_{12}^3}\left(\frac{2 \mathcal{A}}{3}+\frac{3 \mathcal{J}^2}{10 k}+2\mathcal{L}\right)\nonumber\\
&+&\frac{1}{z_{12}^2}\left(\frac{\mathcal{A}'}{3}\pm\frac{\mathcal{J}''}{5}\pm\frac{3 \mathcal{J}'}{5}\pm\frac{\mathcal{J}^3}{20 k^2}\pm\frac{\mathcal{A} \mathcal{J}}{3 k}-\frac{13 \mathcal{G}_- \mathcal{G}_+}{20 k}+\frac{3   \mathcal{J} \mathcal{J}'}{10k}\pm\frac{\mathcal{J}\mathcal{L}}{k}\mp4\mathcal{W}+\mathcal{L}'\right)\nonumber\\
&+&\frac{1}{z_{12}}\left(\frac{\mathcal{A}''}{10}\pm\frac{\mathcal{J}^{(3)}}{20}+\frac{\mathcal{J}^4}{160 k^3}+\frac{\mathcal{A} \mathcal{J}^2}{12 k^2}\mp\frac{\mathcal{G}_- \mathcal{G}_+ \mathcal{J}}{5 k^2}+\frac{\mathcal{J}^2 \mathcal{L}}{4 k^2}\pm\frac{3 \mathcal{J}^2 \mathcal{J}'}{40 k^2}-\frac{3 \mathcal{A}^2}{2 k}\pm\frac{\mathcal{J} \mathcal{A}'}{6 k}
 \right.\nonumber\\
&\pm&\left.
   \frac{\mathcal{A} \mathcal{J}'}{6 k}+\frac{3 \mathcal{A} \mathcal{L}}{5 k}\mp\frac{2 \mathcal{G}_+ \mathcal{S}_-}{k}\pm\frac{2 \mathcal{G}_- \mathcal{S}_+}{k}-\frac{9 \mathcal{G}_\pm
   \mathcal{G}_\mp'}{20 k}-\frac{\mathcal{G}_- \mathcal{G}_+'}{5 k}+\frac{\mathcal{J} \mathcal{J}''}{10 k}\pm\frac{\mathcal{L} \mathcal{J}'}{2 k}+\frac{3 \mathcal{J}'^2}{40 k}
 \right.\nonumber\\
&-&\left.
  \frac{2 \mathcal{J}\mathcal{W}}{k}\pm\frac{\mathcal{J} \mathcal{L}'}{2 k}+\frac{9 \mathcal{L}^2}{10 k}\mp2 \mathcal{W}'+\frac{3 \mathcal{L}''}{10}\right)
\end{eqnarray}
 because some boundary charges do not transform like a primary conformal field, and also there exist some nonlinear terms such as $(\mathcal{JJ})(z)$ ,\, $(\mathcal{G}_{+}\mathcal{G}_{-})(z)$,\,and $(\mathcal{J}\mathcal{G}_{\pm})(z)$, as already discussed in the previous section $(\ref{bhreduction1})$. Therefore, it is required to consider some redefinitions on the boundary charges and gauge parameters as
 \begin{eqnarray}\label{shift11}
\gamma_{-1}\mathcal{L}&\rightarrow&\frac{6}{c}\bigg(\mathcal{L}-\frac{3}{2c} \big(\mathcal{J}\mathcal{J}\big)+{\kappa\over{2}}\mathcal{A}\bigg)
,\,~~\epsilon\rightarrow\epsilon+\frac{\kappa}{2}\bigg(\phi+\frac{6}{c}\upsilon\mathcal{J}\bigg)\\
\vartheta_{-1}\mathcal{A}&\rightarrow&-{9\kappa\over{5c}}\mathcal{A}
,\,~~~~~~~~~~~~~~~~~~~~~~~~~\phi\rightarrow\phi-\frac{3\kappa}{10}\bigg(\phi+\frac{6}{c}\upsilon\mathcal{J}\bigg)\\
\omega_{-2}\mathcal{W}&\rightarrow&{3\kappa\over{5c}}\bigg(\mathcal{W}-\frac{6}{c}\mathcal{JA}\bigg)
,\,~~~~~~~~~~~\upsilon\rightarrow\frac{3\kappa}{10}\upsilon\\
\rho\mathcal{J}&\rightarrow&\frac{3}{c}\mathcal{J}
,\,~~~~~~~~~~~~~~~~~~~~~~~~~~~~\varrho \rightarrow\varrho+\frac{3}{c}\bigg(\epsilon\mathcal{J} +2\upsilon\mathcal{A}\bigg)\\
\sigma_\pm^{-{\frac{1}{2}}}\mathcal{G}_{\pm}&\rightarrow&\mp\frac{3}{c}\mathcal{G}_{\pm}
,\,~~~~~~~~~~~~~~~~~~~~~~~~~\varsigma_{\pm}\rightarrow\pm\varsigma_{\pm}\\
\tau_\pm^{-{\frac{3}{2}}}\mathcal{S}_{\pm}&\rightarrow&\pm\frac{4\kappa}{5c}\mathcal{S}_{\pm}
,\,~~~~~~~~~~~~~~~~~~~~~~\omega_{\pm}\rightarrow\mp\frac{2\kappa}{5}\omega_{\pm}
\end{eqnarray}
where $\kappa=\pm\frac{5i}{2}$,\,which is defined to make a relation with the notation in \cite{Romans:1991wi} at the classical level.

It is important to emphasize that these new variables do not affect the boundary charges. Finally, this leads to operator product expansions of convenient asymptotic symmetry algebra for $\mathcal{N}=(2,2)$ superconformal boundary with a set of conformal generators $\mathcal{G}_{\pm}\rightarrow\mathcal{G}^{+}\pm\mathcal{G}^{-}$ and $\mathcal{S}_{\pm}\rightarrow\mathcal{S}^{+}\pm\mathcal{S}^{-}$ in the complex coordinates by using $(\ref{Qvar2})$.\,
After repeating the same procedure for the barred-sector,\,one can say that the asymptotic symmetry algebra for the loosest set of boundary conditions of $\mathcal{N}=(2,2)$ supergravity  is two copies of the super $\mathcal{W}_3$ algebra with central charge $c\,=\,6k$.\,In this paper,\,we do not explicitly carry out the whole computation to obtain the classical $\mathcal{N}=(2,2)$ super $\mathcal{W}_3$ algebra (see \cite{Romans:1991wi} for the entire quantum  $\mathcal{N}=2$ super $\mathcal{W}_3$ algebra and \cite{Lu:1991ux} for the classical case).\,Recently,\,a detailed derivation of the asymptotic symmetry algebra is also given in \cite{Castro:2020suk} for the $\mathfrak{sl}(3|2)$ case.
\section{Concluding Remarks}

\hspace{0.5cm}
In the present paper,\,it is clearly put forward that the Chern\,-\,Simons formulation of $AdS_3$ (super)gravity also allows a more convenient generalization of higher spin theories for fermionic states as well as for bosonic states. Furthermore, we have also confirmed explicitly that although the higher spin fields do not propagate any degrees of freedom,\,there exists a large class of intriguing nontrivial solutions.\,Specifically, we have built up a candidate solution for $\mathcal{N}=(2,2)$ extended higher spin $AdS_3$ supergravity and scrutinized its asymptotic symmetries.\,

To summarize,\,we have given a brief discussion for $AdS_3$ higher spin supergravity based on Chern\,-\,Simons formulation. We have first worked out $\mathfrak{sl}(2|1) \oplus \mathfrak{sl}(2|1)$ Chern\,-\,Simons $\mathcal{N}=(2,2)$ supergravity theory in detail.\,Then, we have constructed $AdS_3$ higher spin supergravity enlarging $\mathfrak{sl}(2|1) \oplus \mathfrak{sl}(2|1)$ to $\mathfrak{sl}(3|3) \oplus \mathfrak{sl}(3|2)$in the presence of a tower of higher-spin fields up to spin 3.\,Thereafter  we obtained two classical copies of the $\mathfrak{sl}(3|2)_k$ affine algebra on the affine boundary and two copies of super $\mathcal{W}_3$ symmetry algebra on the superconformal boundary as asymptotic symmetry algebras.\,We have also gone through the chemical potentials related to source fields appearing through the temporal components of the connection.\,On the other hand,\,we have seen that Chern\,-\,Simons action is compatible with our boundary conditions,\,has resulted in a finite effect for higher spin fields and a well-defined variational principle. Consequently,\,this method can be considered as a good laboratory for researching the fertile asymptotic structure of extended supergravity.\,It also might be worthwhile to translate our outcomes into the metric formulation language because it will lift our boundary to higher dimensions,\,where is a Chern\,-\,Simons theory.

The results in our paper leave some further investigations which we put in order a few here:

It is a natural question to ask if another class of boundary conditions appearing in literature (see,\,e.g.\,\cite{Compere:2013bya,Afshar:2016wfy,Troessaert:2013fma,Avery:2013dja}),\, whose higher-spin generalization is not as clear as the Grumiller and Riegler 's,\,are consistent with these most general ones,\,and it would be interesting to examine this.\,Besides,\,it has arousing curiosity to get two copies of $\mathcal{N}=(2,2)$ warped superconformal algebras for the supersymmetric boundary conditions of \cite{Troessaert:2013fma} and also to check the supersymmetric extension of  the Avery -Poojary -Suryanarayana boundary conditions \cite{Avery:2013dja,Ozer:2019nkv}.\,In this point,\,it is also worth noting that the boundary conditions are more restrictive than the pure bosonic case \cite{Grumiller:2016pqb}.\,While the first motivation is to extend them,\,it is worth mentioning that these generalizations about boundaries would also have a good potential for novel holographic applications.\,

Finally,\,we close scratching the limits in our debate of the most general boundary conditions of supergravity.\,Many open questions still exist for further investigations,\,e.g.,\,what other boundary conditions from a similar starting point can be attained? Or how can be explained the puzzling result that the related geometries have an entropy?\,Overall,\,we think it is satisfying to see that even in specific instances of three dimensional gravity,\,the asymptotically $AdS$ tale of the most general boundaries recently set forth by Grumiller and Riegler inspires new and unexpected innovations.\,

Last but not least,\,according to \cite{Grumiller:2016pqb} and \cite{Ozer:2019nkv} there is $\mathcal{N}=(2,2)$ extended supergravity with new boundaries.\,However,\,it is an open problem to decide whether we end up to its enough higher order $\mathcal{N}$ extension by taking the Grumiller-Riegler method as we perform in this paper.\,Therefore,\,our results presented in this paper can be extended in various ways. One possible extension is $\mathcal{N}=3$ supergravity theory in $AdS_3$. In this context,\,the details of this possible extension will be examined in our forthcoming paper.
\section{Acknowledgments}
This work was supparted by Istanbul Technical University Scientific Research Projects Department(ITU BAP,project number:40199).


\end{document}